\documentclass[aps,pre,twocolumn,floats,groupedaddress]{revtex4-1}
\usepackage{graphicx}
\usepackage{verbatim}
\usepackage{amsmath}
\usepackage{float}

\def\lb{\lambda_B}

\def\etal{{\it et al.}}

\begin{document}


\title{Structure and Stability of Charged Colloid-Nanoparticle Mixtures}


\author{Braden M. Weight and Alan R. Denton}
\email[]{alan.denton@ndsu.edu}
\affiliation{Department of Physics, North Dakota State University, Fargo, 
ND, U.S.A. 58108-6050}


\date{\today}

\begin{abstract}
Physical properties of colloidal materials can be modified by addition of nanoparticles.
Within a model of like-charged mixtures of particles governed by effective electrostatic 
interactions, we explore the influence of charged nanoparticles on the structure and 
thermodynamic phase stability of charge-stabilized colloidal suspensions.  Focusing on 
salt-free mixtures of particles of high size and charge asymmetry, interacting via 
repulsive Yukawa effective pair potentials, we perform molecular dynamics simulations 
and compute radial distribution functions and static structure factors.  Analysis of 
these structural properties indicates that increasing the charge and concentration of 
nanoparticles progressively weakens correlations between charged colloids.  We show that
addition of charged nanoparticles to a suspension of like-charged colloids can induce 
a colloidal crystal to melt and can facilitate aggregation of a fluid suspension due 
to attractive van der Waals interactions.  We attribute the destabilizing influence 
of charged nanoparticles to enhanced screening of electrostatic interactions, which 
weakens repulsion between charged colloids.  This interpretation is consistent with recent 
predictions of an effective interaction theory of charged colloid-nanoparticle mixtures. 
\end{abstract}


\maketitle


\section{Introduction}\label{Introduction}
Suspensions of colloidal (nm-$\mu$m-sized) particles, dispersed in a solvent by 
Brownian motion, are characterized by extreme sensitivity of thermal 
and mechanical properties to environmental conditions~\cite{Pusey1991}.
Relatively small changes in internal composition or external fields can modify 
interparticle interactions and profoundly affect thermodynamic stability and phase behavior 
of these soft materials.   Such diversity and tunability of materials properties makes 
colloidal suspensions appealing for many practical applications, e.g., in the food, 
pharmaceutical, and consumer products industries~\cite{Hamley2000,jones2002}.

Opposing thermodynamic stability of colloidal suspensions are attractive van der Waals 
interparticle forces~\cite{Israelachvili1992}, which drive particles to aggregate and 
can lead to coagulation and sedimentation.  Colloidal particles can be stabilized against 
aggregation by introducing either steric forces, e.g., by adsorbing or grafting polymers 
or surfactants to their surfaces, or electrostatic forces, by creating surface
charges via dissociation of counterions.  Steric and charge stabilization are 
two established mechanisms known to prevent aggregation~\cite{Pusey1991,Evans1999,
Hamley2000,jones2002,Israelachvili1992}.

Addition of nanoparticles can also affect structure and thermodynamic stability of
colloidal suspensions.  In a series of experiments, Lewis \etal~\cite{Lewis-2001-pnas,
Lewis-2001-langmuir,Lewis-2005-langmuir,chan-lewis2008-langmuir,Lewis-2008-langmuir} 
observed that addition of charged zirconia or polystyrene nanoparticles to aqueous 
suspensions of weakly charged silica colloidal microspheres affected the tendency of 
the colloidal particles to aggregate.  
With increasing nanoparticle concentration, suspensions transitioned from unstable 
to stable and back to unstable.  Lewis \etal~postulated that nanoparticles may inhibit 
aggregation by forming halos around the colloidal particles.  Numerous independent
experimental studies~\cite{weeks-luijten-lewis2005,weeks-lewis2005,willing2009,
buzzaccaro-piazza-parola2010,wunder2011,ngai2012,walz-langmuir2013,walz-langmuir2015,
Kazi2015,herman2015,zubir2015} have confirmed that nanoparticle haloing stabilizes
mixtures of weakly charged colloids and strongly charged nanoparticles that can adsorb 
to the colloidal surfaces.  At intermediate concentrations, adsorbed nanoparticles  
increase the magnitude of the zeta potential of the colloids, promoting charge 
stabilization, while at higher concentrations, nanoparticles screen repulsive 
electrostatic interactions, destabilizing the suspension.  The role of nanoparticles
in screening electrostatic interactions is not, however, well understood.

Charged colloidal mixtures have been modeled by a variety of theoretical and 
computational methods, including computer simulations
~\cite{louis-allahyarov-loewen-roth2002,Luijten2004,Linse2005,Dijkstra2007,Dijkstra2010},
integral-equation theory~\cite{Klein1991-jpcm,Klein1992-jpcm,Lowen1991-jpcm,Naegele1990,
Klein1992-pra,Louis2004,schweizer2008,Medina-Noyola2010,Medina-Noyola2011,
Moncho-Jorda-SM2018}, and Poisson-Boltzmann theory~\cite{Torres2008-pre,Torres2008-jcp,
castaneda-priego2011,ruckenstein-jpcb2013,ruckenstein-csa2013}.  Several recent studies 
of highly asymmetric mixtures of charged colloids and nanoparticles~\cite{Luijten2004,
Louis2004,schweizer2008,ruckenstein-jpcb2013,ruckenstein-csa2013} yielded effective 
colloid-colloid pair interactions that support the postulated nanoparticle haloing 
mechanism for certain system parameters.  However, while previous studies have focused 
on weakly charged colloids mixed with oppositely-charged nanoparticles, relatively 
little is known about the influence on suspensions of strongly charged colloids of 
like-charged nanoparticles. 

Motivated by previous experimental observations and modeling results, we conducted 
a computer simulation study to explore the impacts of nanoparticle doping on 
the structure and phase stability of charge-stabilized colloidal suspensions.  
Our study is distinguished from previous work by focusing on mixtures in which 
both particle species carry significant charge of the same sign.  To highlight 
the influence of electrostatic interparticle interactions, we adopt a practical 
coarse-grained model in which charged colloids and nanoparticles interact via 
physically consistent effective pair potentials.

In Sec.~\ref{Model}, we define our model, which is derived from the primitive model 
of charged colloid-nanoparticle mixtures  by averaging over microion degrees of freedom.
In Sec.~\ref{Methods}, we outline our methods of molecular dynamics simulation and 
analysis of structural properties.  Section~\ref{Results} presents numerical results 
for radial distribution functions and static structure factors of model salt-free mixtures 
over ranges of system parameters, including nanoparticle charge and concentration.
From these results, we identify general trends in the influence of nanoparticles 
on pair correlations between charged colloids and on the structure and phase stability 
of bulk suspensions.  Finally, in Sec.~\ref{Conclusions}, we summarize and conclude 
with an outlook for future work.

\section{Coarse-Grained Model of Colloid-Nanoparticle Mixtures}\label{Model}
\begin{figure}[t]
\includegraphics[width=0.9\columnwidth,angle=0]{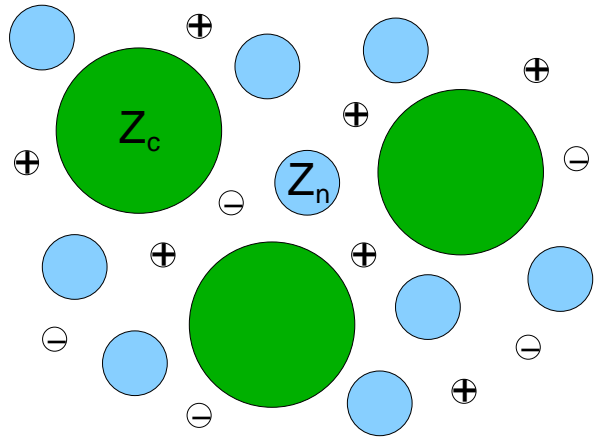}
\caption{Primitive model of colloid-nanoparticle mixture:
charged colloids (valence $Z_c$), nanoparticles (valence $Z_n$), 
and microions (counterions and coions) in an implicit solvent.
\label{modelCartoon}
}
\end{figure}

We consider a bidisperse mixture of $N_c$ colloidal particles and $N_n$ nanoparticles,
of average number densities $n_c=N_c/V$ and $n_n=N_n/V$, dispersed in a solvent 
in a volume $V$ at absolute temperature $T$, along with $N_{\mu}$ microions: 
$N_+$ counterions and $N_-$ coions (see Fig.~\ref{modelCartoon}).  The colloids and 
nanoparticles are modeled as charged hard spheres (macroions), of respective radii 
$a_c$ and $a_n$ and valences $Z_c$ and $Z_n$, and the microions as point charges of 
valences $\pm z$.  For simplicity, we consider a symmetric electrolyte of salt 
ion pairs with the same valence as the counterions.  The model is easily generalized,
however, to describe asymmetric electrolytes and finite-sized microions.  The surface 
charges of the colloids and nanoparticles originate from dissociation of counterions.
Assuming like-charged (negative) particles, global electroneutrality dictates that 
\begin{equation}
Z_cN_c+Z_nN_n=z(N_+-N_-)~.
\label{electroneutrality}
\end{equation}

As an underlying explicit model, we take the primitive model of charged colloidal
mixtures, in which the solvent is idealized as a continuum medium of dielectric 
constant $\epsilon$.  Across this effective medium, the van der Waals interaction 
between particles $i$ and $j$, of radii $a_i$ and $a_j$, at separation $r$ between
the particle centers, is approximated within the Hamaker theory~\cite{Israelachvili1992}
by an effective pair potential of the form
\begin{eqnarray}
v_{\rm \scriptscriptstyle vdW}(r)&=&-\frac{A_{ij}}{6}\left[\frac{2a_i a_j}{r^2-(a_i+a_j)^2}
+\frac{2a_i a_j}{r^2-(a_i-a_j)^2}\right.
\nonumber\\[1ex]
&+&\left.\ln\left(\frac{r^2-(a_i+a_j)^2}{r^2-(a_i-a_j)^2}\right)\right],
\quad r>a_i+a_j,
\label{vdW}
\end{eqnarray}
where the Hamaker constant $A_{ij}$ depends on the dielectric constants of both the 
particles and the medium.  Since we restrict attention here to like-charged particles 
whose repulsive electrostatic interactions are sufficiently strong to prevent nanoparticle 
adsorption, we neglect dielectric polarization effects, as in previous studies of charged 
colloidal mixtures~\cite{Luijten2004,Linse2005,Dijkstra2007,Dijkstra2010,Klein1991-jpcm,
Klein1992-jpcm,Lowen1991-jpcm,Naegele1990,Klein1992-pra,Louis2004,
schweizer2008,Medina-Noyola2010,Medina-Noyola2011, Torres2008-pre,Torres2008-jcp,
castaneda-priego2011,ruckenstein-jpcb2013,ruckenstein-csa2013}.  Although induced 
polarization charges may affect self-assembly of strongly coupled, oppositely 
charged colloidal mixtures~\cite{Luijten2014}, charge-charge interactions are 
expected to dominate in the moderately coupled, like-charged systems investigated here.
Furthermore, our simulations are for the most part limited to system parameters for which, 
as shown below, electrostatic interactions dominate over van der Waals interactions.
The systems modeled here thus differ importantly from experimental systems in which 
van der Waals attraction between nanoparticles and weakly or oppositely charged 
colloidal particles leads to nanoparticle adsorption and haloing~\cite{Lewis-2001-pnas,
Lewis-2001-langmuir,Lewis-2005-langmuir,chan-lewis2008-langmuir,Lewis-2008-langmuir}.

Within the primitive model, all charged particles interact via electrostatic 
(Coulomb) pair potentials.  The model system is formally governed by a Hamiltonian,
$H = H_0 + H_{\rm el}$, which naturally separates into a reference term $H_0$, 
including the total kinetic energy and particle hard-core interactions,
and a term $H_{\rm el}$ that represents the total electrostatic energy:
\begin{equation}
H_{\rm el} = H_{mm} + H_{\mu\mu} + H_{m\mu}~.
\label{Hel}
\end{equation} 
The three terms on the right side of Eq.~(\ref{Hel}) account for pair interactions 
among colloidal and nanoparticle macroions ($m$) and microions ($\mu$).  These terms 
can be expressed more explicitly as
\begin{equation}
H_{mm} = \sum_{i<j=1}^{N_c}v_{cc}(r_{ij}) + \sum_{i<j=1}^{N_n}v_{nn}(r_{ij})
+ \sum_{i=1}^{N_c}\sum_{j=1}^{N_n}v_{cn}(r_{ij})~,
\label{Hm}
\end{equation}
\begin{equation}
H_{\mu\mu} = \sum_{i<j=1}^{N_+}v_{++}(r_{ij}) + \sum_{i<j=1}^{N_-}v_{--}(r_{ij})
+ \sum_{i=1}^{N_+}\sum_{j=1}^{N_-}v_{+-}(r_{ij})~,
\label{Hmu}
\end{equation}
\begin{equation}
H_{m\mu} = \sum_{i=1}^{N_c}\sum_{j=1}^{N_{\mu}}v_{c\mu}(r_{ij})
+ \sum_{i=1}^{N_n}\sum_{j=1}^{N_{\mu}}v_{n\mu}(r_{ij})~,
\label{Hmmu}
\end{equation}
where $N_{\alpha}$ denotes the number of particles of species ${\alpha}$ and
$v_{\alpha\beta}(r_{ij})$ is the Coulomb pair potential between particle $i$
of species $\alpha$ and particle $j$ of species $\beta$ separated by 
center-to-center distance $r_{ij}$.

The primitive model of charged colloidal mixtures [Eqs.~(\ref{Hm})-(\ref{Hmmu})],
wherein all particles interact via long-range Coulomb potentials, 
poses considerable challenges for computational modeling.  Recently, 
Chung and Denton~\cite{Chung-Denton2013} developed a general effective interaction 
theory for polydisperse suspensions of charged colloids.  Extending to mixtures 
an established coarse-graining approach~\cite{vanRoij1997,Denton1999,Denton2000,DentonBook}, 
which traces out microion degrees of freedom from the partition function, the theory maps 
the multicomponent macroion-microion mixture onto a system of only pseudo-macroions 
governed by effective electrostatic interactions.  Within a linear approximation 
for the response of microion densities to macroion potentials and a 
random phase approximation for microion response functions, the colloidal and 
nanoparticle macroions interact via repulsive hard-core Yukawa (screened-Coulomb)
effective pair potentials.  In thermal ($k_BT$) units, the effective 
electrostatic pair potentials take the form
\begin{equation}
v_{{\rm\scriptscriptstyle Y}}(r)=Z_i Z_j\lb\frac{e^{\kappa(a_i+a_j)}}{(1+\kappa a_i)(1+\kappa a_j)}~
\frac{\displaystyle e^{-\kappa r}}{\displaystyle r}, 
\quad r>a_i+a_j,
\label{vrY}
\end{equation}
where 
\begin{equation}
\kappa=\sqrt{\frac{\displaystyle 4\pi z^2\lambda_B(|Z_c|n_c+ |Z_n|n_n+2n_s)}
{\displaystyle 1-\phi}}
\label{kappa}
\end{equation}
is the Debye screening constant, $\lambda_B=e^2/(\epsilon k_BT)$ is the Bjerrum length, 
$k_B$ is the Boltzmann constant, and $\phi=(4\pi/3)(n_ca_c^3+n_na_n^3)$ is the fraction 
of volume from which the microions are excluded by the particle hard cores.
Similar effective pair potentials have been derived previously by
integral-equation methods~\cite{Naegele1990} and adopted in simulation 
studies~\cite{Allahyarov1998,Allahyarov2009,Hynninen2006}, albeit neglecting
the excluded-volume factor of $1/(1-\phi)$ in the screening constant, 
which can be significant in concentrated suspensions.
The effective interaction theory also predicts a one-body volume energy $E_0$,
which accounts for the microion entropy and macroion-microion interaction 
energy~\cite{Chung-Denton2013}.
While not affecting structural properties, the volume energy influences 
thermodynamic properties through its dependence on average macroion densities,
which can be especially significant in deionized suspensions.
The coarse-grained model is thus governed by an effective Hamiltonian,
$H_{\rm eff}=H_0+E_0+H_{\rm el, eff}$, where the last term,
\begin{eqnarray}
H_{\rm el, eff}&=&\sum_{i<j=1}^{N_c}v_{cc, \rm\scriptscriptstyle Y}(r_{ij}) + 
\sum_{i<j=1}^{N_n}v_{nn, \rm \scriptscriptstyle Y}(r_{ij})
\nonumber\\[1ex]
&+&\sum_{i=1}^{N_c}\sum_{j=1}^{N_n}v_{cn, \rm \scriptscriptstyle Y}(r_{ij}),
\label{Heff}
\end{eqnarray}
includes all electrostatic effective pair potentials.  Compared with the primitive model 
of charged colloids, the coarse-grained model greatly facilitates computational modeling 
by vastly reducing the number of particles, the microions being included only implicitly
through $\kappa$, and by replacing long-range Coulomb pair potentials with 
shorter-range Yukawa potentials.

Combining the van der Waals pair potential [Eq.~(\ref{vdW})] with the electrostatic 
effective pair potential [Eq.~(\ref{vrY})] yields the classic Derjaguin-Landau-Verwey-Overbeek (DLVO)
pair potential~\cite{Pusey1991,Hamley2000,jones2002,Israelachvili1992,Evans1999,DL1941,VO1948} 
for charged colloidal mixtures:  
\begin{equation}
v_{\rm \scriptscriptstyle DLVO}(r)=v_{\rm \scriptscriptstyle vdW}(r)+v_{\rm \scriptscriptstyle Y}(r).
\label{dlvo}
\end{equation}
The negative divergence of the van der Waals potential near contact of the 
particle hard cores renders uncharged or weakly charged colloids and nanoparticles
prone to irreversible aggregation.  If the particles are sufficiently charged, however, 
mutual electrostatic repulsion can stabilize a suspension against aggregation.

\begin{figure}
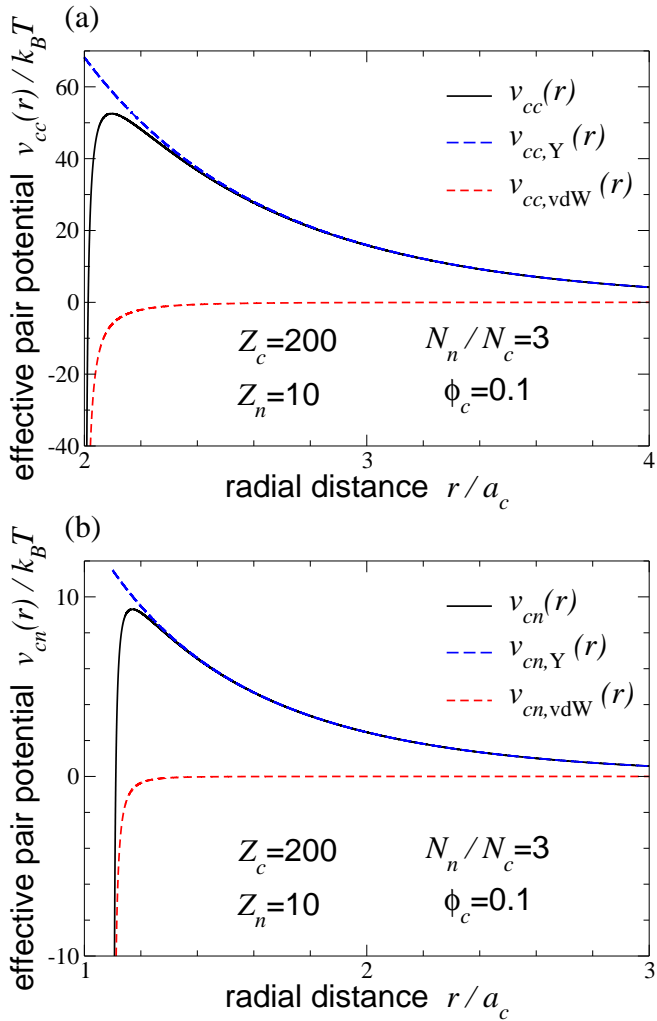

\includegraphics[width=0.48\textwidth]{vrcc.zc200.eps}
\includegraphics[width=0.48\textwidth]{vrcn.zc200.eps}
\caption{Effective pair potentials ($k_BT$ units) in deionized (salt-free) mixtures of charged colloids 
and nanoparticles.  (a) Colloid-colloid and (b) colloid-nanoparticle effective pair potentials 
for colloid radius $a_c=50$ nm and valence $Z_c=200$;
nanoparticle radius $a_n=5$ nm and valence $Z_n=10$;
colloid volume fraction $\phi_c=$ 0.1, number of nanoparticles per colloid $N_n/N_c=3$,
and Hamaker constants $A_{cc}=A_{cn}=5\times 10^{-20}$ J.
The total effective pair potential (solid curves) is the sum of the electrostatic Yukawa 
potential (dashed curves) and the van der Waals potential (short-dashed curves).
\label{vr1}}
\end{figure}

\begin{figure}
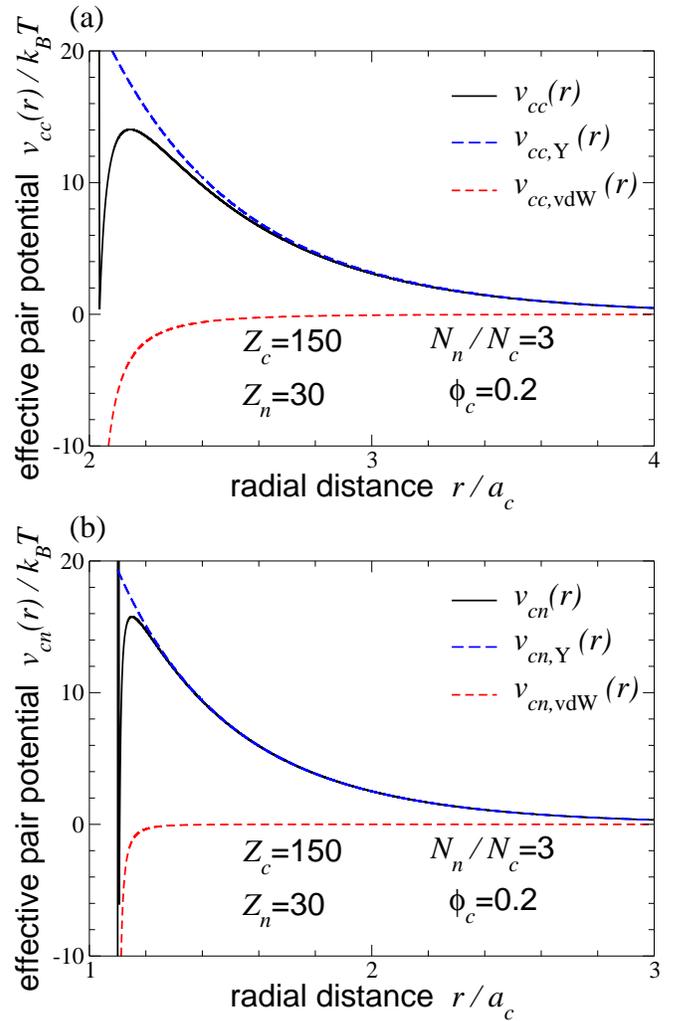

\includegraphics[width=0.48\textwidth]{vrcc.zc150.zn30.eps}
\includegraphics[width=0.48\textwidth]{vrcn.zc150.zn30.eps}
\caption{Same as Fig.~\ref{vr1}, but for more weakly charged colloids ($Z_c=150$), more 
strongly charged nanoparticles ($Z_n=30$), and higher colloid volume fraction ($\phi_c=0.2$).
Steric interactions, modeled by short-range, repulsive pair potentials [Eq.~(\ref{LJ})],
are added to cut off the van der Waals divergence at contact of the particle hard cores.  
Note the changes of vertical scale.
\label{vr2}}
\end{figure}

Figures~\ref{vr1} and \ref{vr2} illustrate the relative contributions of the van der Waals 
and Yukawa terms to the colloid-colloid and colloid-nanoparticle effective pair potentials 
for two different combinations of particle valences.  To accentuate the influence of 
van der Waals interactions, we have here chosen unusually large Hamaker constants,
$A_{cc}=A_{cn}=5\times 10^{-20}$ J, more typical of metallic particles~\cite{Israelachvili1992}.
In the case of relatively strongly charged colloids and weakly charged nanoparticles (Fig.~\ref{vr1}), 
electrostatic interactions dominate van der Waals interactions, except for very near 
contact.  The total colloid-colloid effective pair potential then has a barrier height 
far exceeding the typical thermal energy ($k_BT$).  In contrast, for more weakly
charged colloids and strongly charged nanoparticles (Fig.~\ref{vr2}), electrostatic 
and van der Waals interactions compete more evenly near contact, resulting in 
a lower barrier height of $v_{cc}(r)$.

While most of the simulations described in Sec.~\ref{Results} were performed for 
systems governed by pair potentials typified by Fig.~\ref{vr1}, where van der Waals 
interactions can be neglected, we also simulated a system governed by the pair potentials 
depicted in Fig.~\ref{vr2}.  In the latter case, where particles interact via the full 
DLVO pair potential, it was necessary to cut off the divergence of the van der Waals
potential near contact of the particle hard cores.  This cutting off was achieved by 
including short-range steric interactions, modeled by shifted, repulsive Lennard-Jones 
pair potentials of the form 
\begin{equation}
v_{\rm \scriptscriptstyle LJ}(r)=
\left\{ \begin{array}
{l@{\quad}l}
4\epsilon_{\rm \scriptscriptstyle LJ}\left[\left(\frac{\displaystyle\sigma_{\rm \scriptscriptstyle LJ}}
{\displaystyle r}\right)^{12}-\left(\frac{\displaystyle\sigma_{\rm \scriptscriptstyle LJ}}
{\displaystyle r}\right)^6\right]+\epsilon_{\rm \scriptscriptstyle LJ}~,
& r<r_c \\[2ex]
0~,
& r>r_c,
\end{array} \right.
\label{LJ}
\end{equation}
where $r_c=2^{\scriptscriptstyle 1/6}\sigma_{\rm \scriptscriptstyle LJ}$ and 
$\sigma_{\rm \scriptscriptstyle LJ}$ and $\epsilon_{\rm \scriptscriptstyle LJ}$ 
are chosen to yield a total pair potential that is steeply repulsive near core
contact and has a well depth of at least 10 $k_BT$.
The required values of $\epsilon_{\rm \scriptscriptstyle LJ}$ are in the range
$10^6-10^7~k_B T$, closely mimicking a hard-sphere pair potential.
In all cases considered here, the colloid-nanoparticle effective pair potential is 
electrostatically dominated, the barrier height of $v_{cn}(r)$ being $\gg k_BT$,
precluding adorption of nanoparticles onto colloids.  

\begin{figure}[t]
\includegraphics[width=0.5\textwidth]{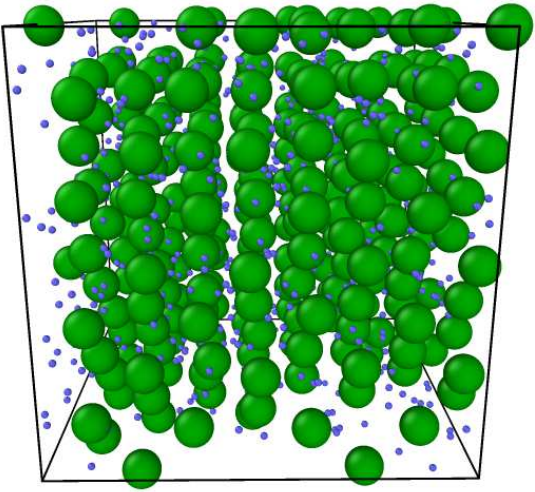}
\caption{Snapshot from molecular dynamics simulation of coarse-grained model of a 
colloid-nanoparticle mixture.  Colloids (nanoparticles) are depicted as green 
(blue) spheres.
\label{snapshot}
}
\end{figure}
\section{Computational Methods}\label{Methods}
For the coarse-grained model of charged colloid-nanoparticle mixtures described 
by Eqs.~(\ref{vrY})-(\ref{LJ}), we performed molecular dynamics (MD) simulations, 
using LAMMPS~\cite{lammps,plimpton1995} to integrate (by Verlet's method) Newton's 
equations of motion.  All of our runs were for like-charged (i.e., mutually repulsive)
particles.  In most of the runs, the particle valences were chosen sufficiently 
high that electrostatic interactions dominated the effective pair potentials, 
as exemplified by Fig.~\ref{vr1}.  In these cases, since the potential barrier was 
so high ($\gg k_BT$), we were able to neglect van der Waals and steric interactions
and use a time step of 10 fs.  In a few runs, detailed in Sec.~\ref{Results}, 
we set the colloid valence low enough that van der Waals and steric interactions 
between the particle hard cores had to be included (see Fig.~\ref{vr2}).

The effective pair potentials were truncated at half the box length, ensuring that
the cut-off distance exceeded $10/\kappa$ (i.e., 10 screening lengths) for most runs.  
The simulations were conducted in the canonical ensemble for fixed numbers of 
particles in a cubic box of fixed volume, subject to periodic boundary conditions,
at an average temperature of $T=$ 293 K, maintained by a Nos\'e-Hoover thermostat.  
While this protocol proves efficient for computing equilibrium properties, other methods 
(e.g., Langevin dynamics) may be more appropriate for studying dynamical properties.
To model aqueous suspensions, we set $\lambda_B=$ 0.714 nm, corresponding to 
$\epsilon\simeq 80$ for water.  Variation of the nanoparticle concentration was 
facilitated by initializing all particles on the sites of a cubic lattice with a basis 
of up to 16 atoms.  After an annealing stage of $10^5$ time steps, during which the 
temperature was steadily ramped down from 1000 K to 293 K, followed by an equilibration 
stage of $10^5$ steps, we computed structural quantities by averaging over particle 
trajectories for an additional $10^6$ time steps.

From particle configurations, we computed the partial radial distribution functions,
\begin{equation}
g_{\alpha \beta}(r) = \frac{V}{x_{\alpha}x_{\beta}N^2}
\sum\limits_{i=1}^{N_{\alpha}}\sideset{}{'}\sum_{j=1}^{N_{\beta}}
\langle\delta({\bf r}+{\bf r}_j-{\bf r}_i)\rangle~,
\label{gr}
\end{equation}
and the partial static structure factors~\cite{HansenMcDonald}, 
\begin{equation}
S_{\alpha \beta}(q) = x_{\alpha}\delta_{\alpha\beta} + \frac{1}{N}\sum_{i=1}^{N_{\alpha}}
\sideset{}{'}\sum_{j=1}^{N_{\beta}}\left\langle\frac{\sin(qr_{ij})}{qr_{ij}}\right\rangle~, 
\label{Sq}
\end{equation}
where $q$ is wave vector magnitude, $x_{\alpha}=N_{\alpha}/N$ is the concentration 
of species $\alpha$, $\delta({\bf r})$ is the Dirac delta function, $\delta_{\alpha\beta}$ 
is the Kronecker delta function, the prime on the sum implies exclusion of self-interactions,
and angular brackets represent a time average.  In computing averages, we sampled
independent configurations separated by intervals of $10^3$ time steps, thus including
a total of $10^3$ configurations.  As we are particularly interested in the influence 
of nanoparticles on the colloid-colloid structure factor,
\begin{equation}
S_{cc}(q) = x_c + \frac{1}{N}\sum_{i<j=1}^{N_c}
\left\langle\frac{\sin(qr_{ij})}{qr_{ij}}\right\rangle~,
\label{Sccq}
\end{equation}
we also define the scaled function,
\begin{equation}
\frac{S_{cc}(q)}{x_c} = 1 + \frac{1}{N_c}\sum_{i<j=1}^{N_c}
\left\langle\frac{\sin(qr_{ij})}{qr_{ij}}\right\rangle~,
\label{Sccq-scaled}
\end{equation}
for direct comparison with the structure factor of the corresponding nanoparticle-free suspension. 
To test and validate our structural analysis methods, we reproduced previously reported
static structure factors of binary Lennard-Jones liquid mixtures~\cite{kob-andersen1995}.

\section{Results and Discussion}\label{Results}
To explore the influence of charged nanoparticles on structural properties of 
charge-stabilized colloidal suspensions, we performed a series of simulations 
over ranges of nanoparticle charge and concentration.  Taking care to restrict 
system parameters to ranges in which the Yukawa effective pair potentials 
have been validated, we considered only particle valences below established charge 
renormalization thresholds~\cite{Denton2008,Denton2010,Lu-Denton2010}, 
$Z_c\lambda_B/a_c\lesssim 6$ and $Z_n\lambda_B/a_n\lesssim 6$. 
To reduce the parameter space of these complex systems, we set $N_c=500$ and $Z_c=200$, 
fixed the particle radii at $a_c=$ 50 nm and $a_n=$ 5 nm, set $z=1$ (monovalent microions) 
to ensure weak electrostatic coupling ($z^2\lambda_B/a_c=0.014$, $z^2\lambda_B/a_n=0.14$), 
and considered only salt-free suspensions ($n_s=0$).  We performed runs for colloid 
volume fractions $\phi_c=(4\pi/3)n_c a_c^3$ ranging from 0.01 to 0.2 and checked that 
finite-size effects were negligible by repeating runs for $N_c=256$.  
Figure~\ref{snapshot} shows a snapshot from a typical simulation.

\begin{figure}
\includegraphics[width=0.48\textwidth]{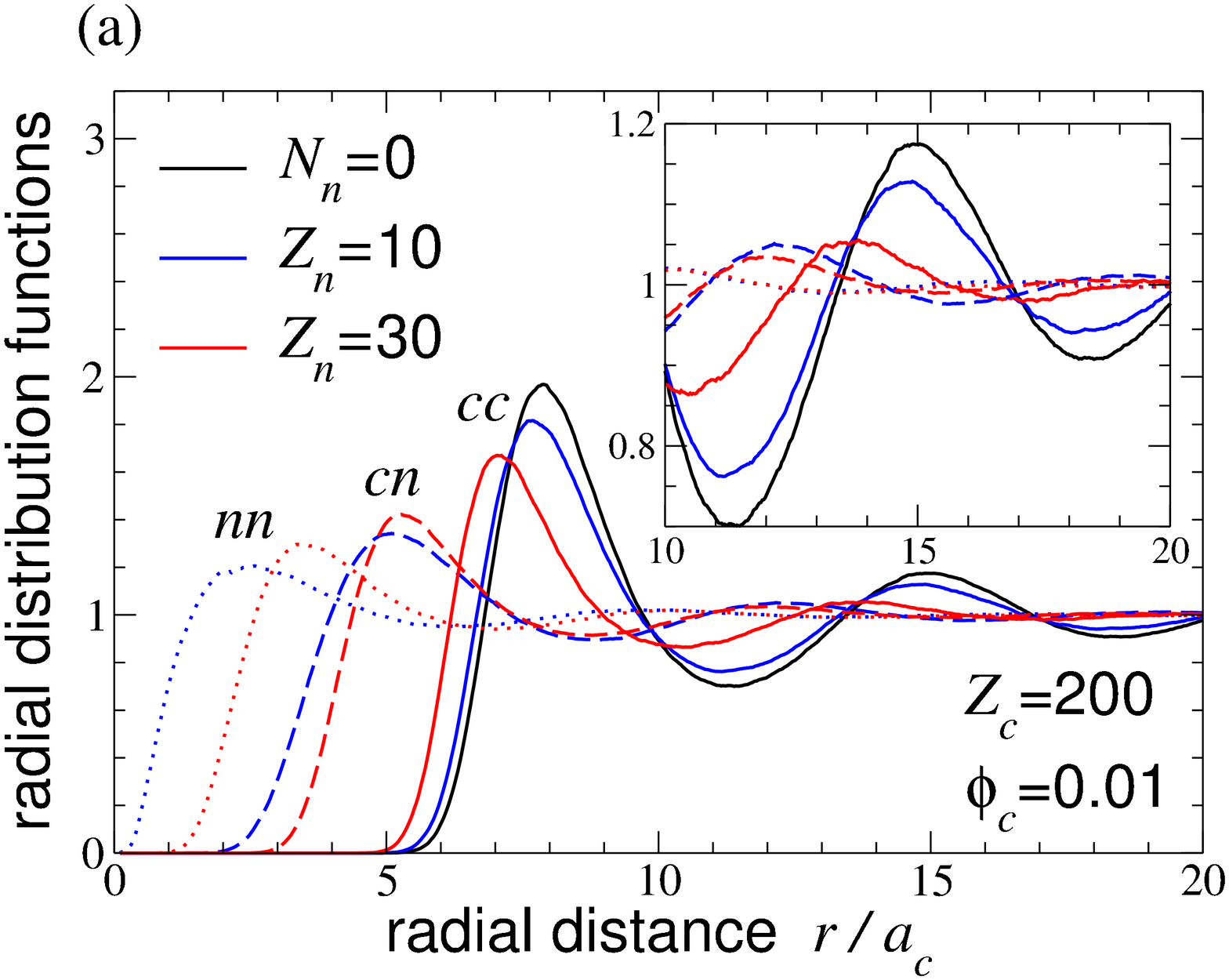}
\includegraphics[width=0.48\textwidth]{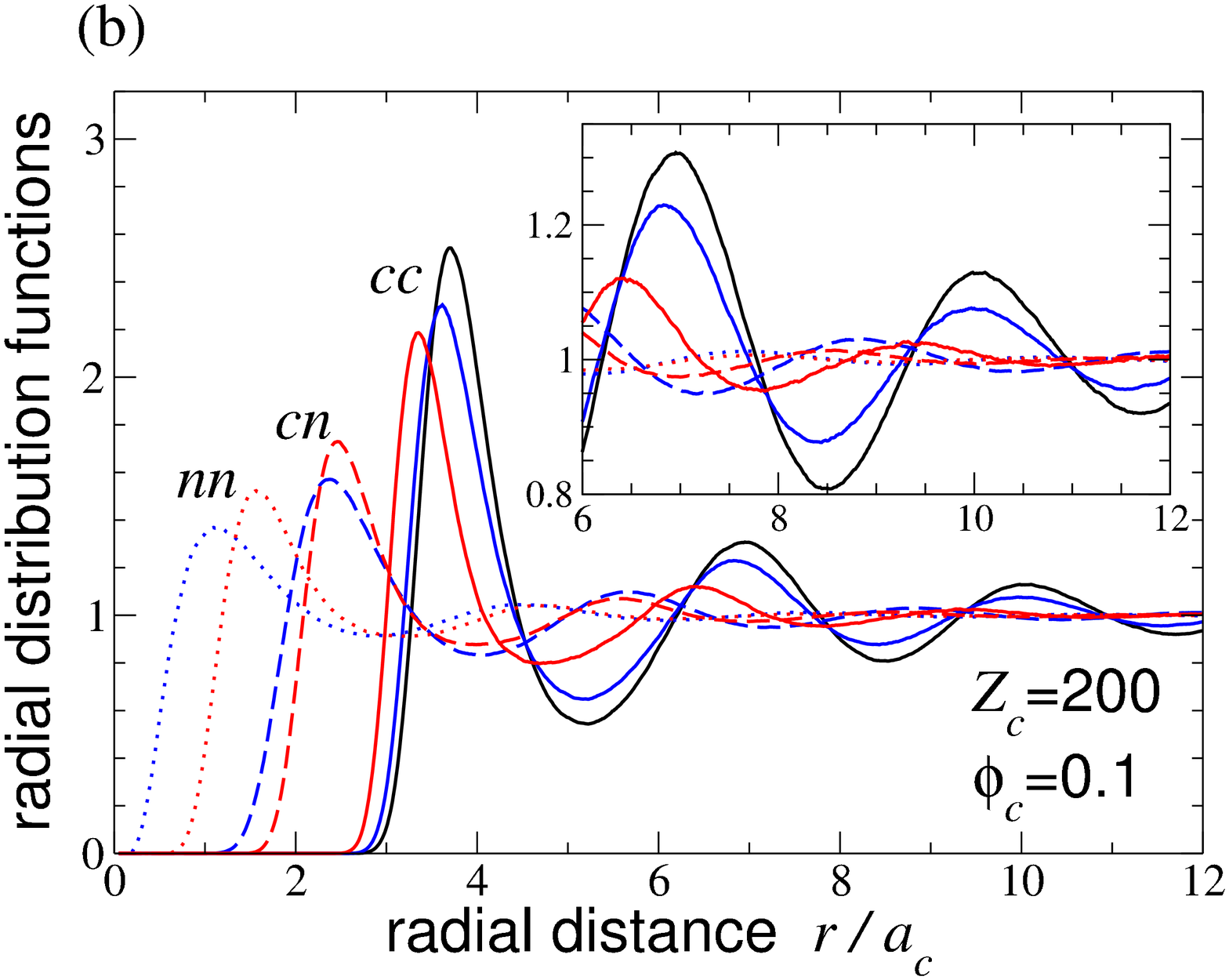}
\includegraphics[width=0.48\textwidth]{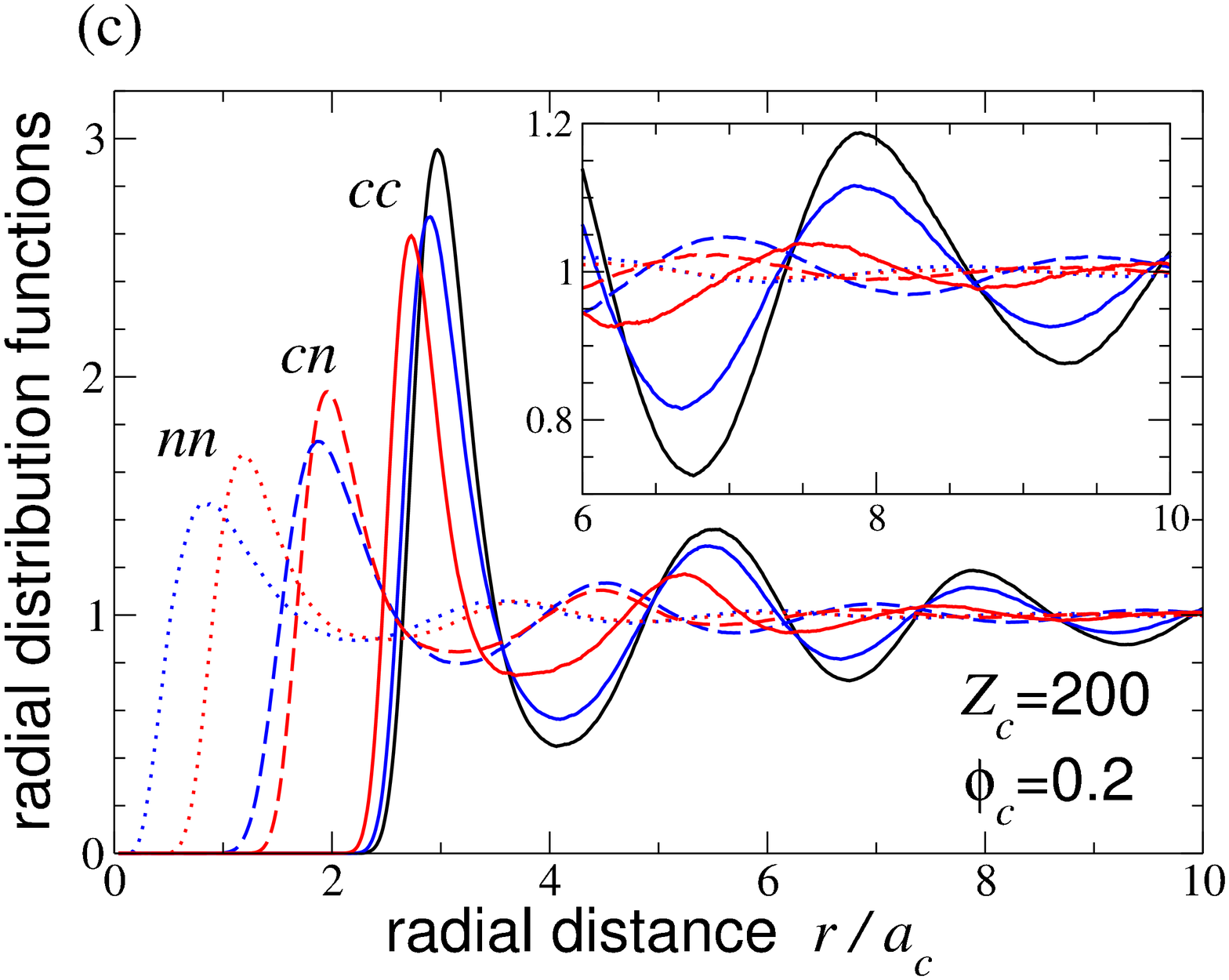}
\caption{Radial distribution functions from MD simulations of model salt-free 
mixtures of like-charged colloids and nanoparticles, interacting via effective 
pair potentials typical of those in Fig.~\ref{vr1}.
Results are shown for colloid number $N_c=500$, radius $a_c=50$ nm, and valence $Z_c=200$;
nanoparticle number $N_n=1500$, radius $a_n=5$ nm, and valences $Z_n=10$ (blue curves) 
and 30 (red curves); and colloid volume fractions $\phi_c=0.01$ (a), 0.1 (b), and 0.2 (c).
For comparison, $g_{cc}(r)$ of nanoparticle-free ($N_n=0$) suspensions are also shown 
(black curves).  Solid, dashed, and dotted curves, whose main peaks are labeled 
$cc$, $cn$, and $nn$, represent $g_{cc}(r)$, $g_{cn}(r)$, and $g_{nn}(r)$, respectively.
Insets magnify long-ranged behavior.
\label{rdf1}}
\end{figure}
\begin{figure}
\includegraphics[width=0.48\textwidth]{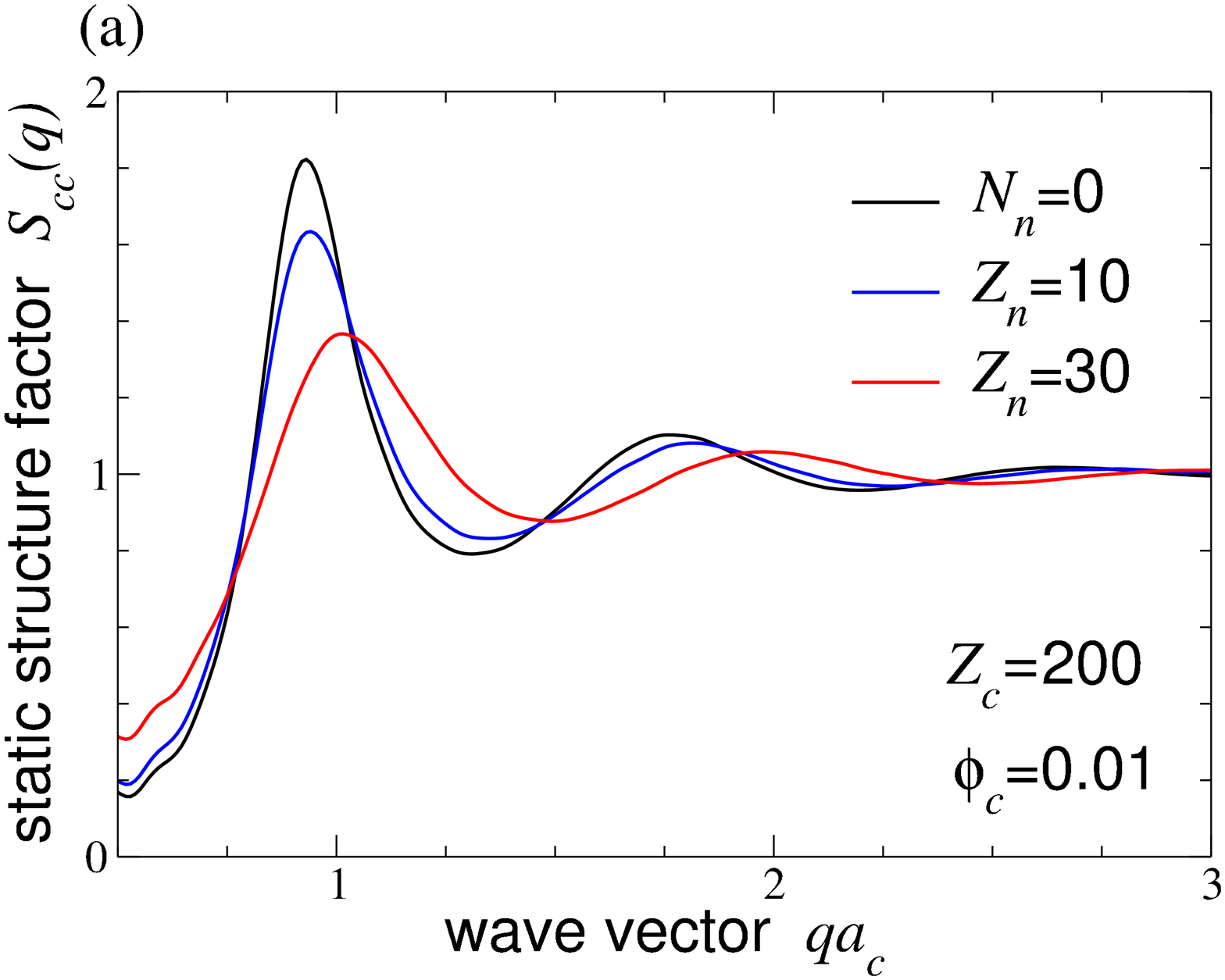}
\includegraphics[width=0.48\textwidth]{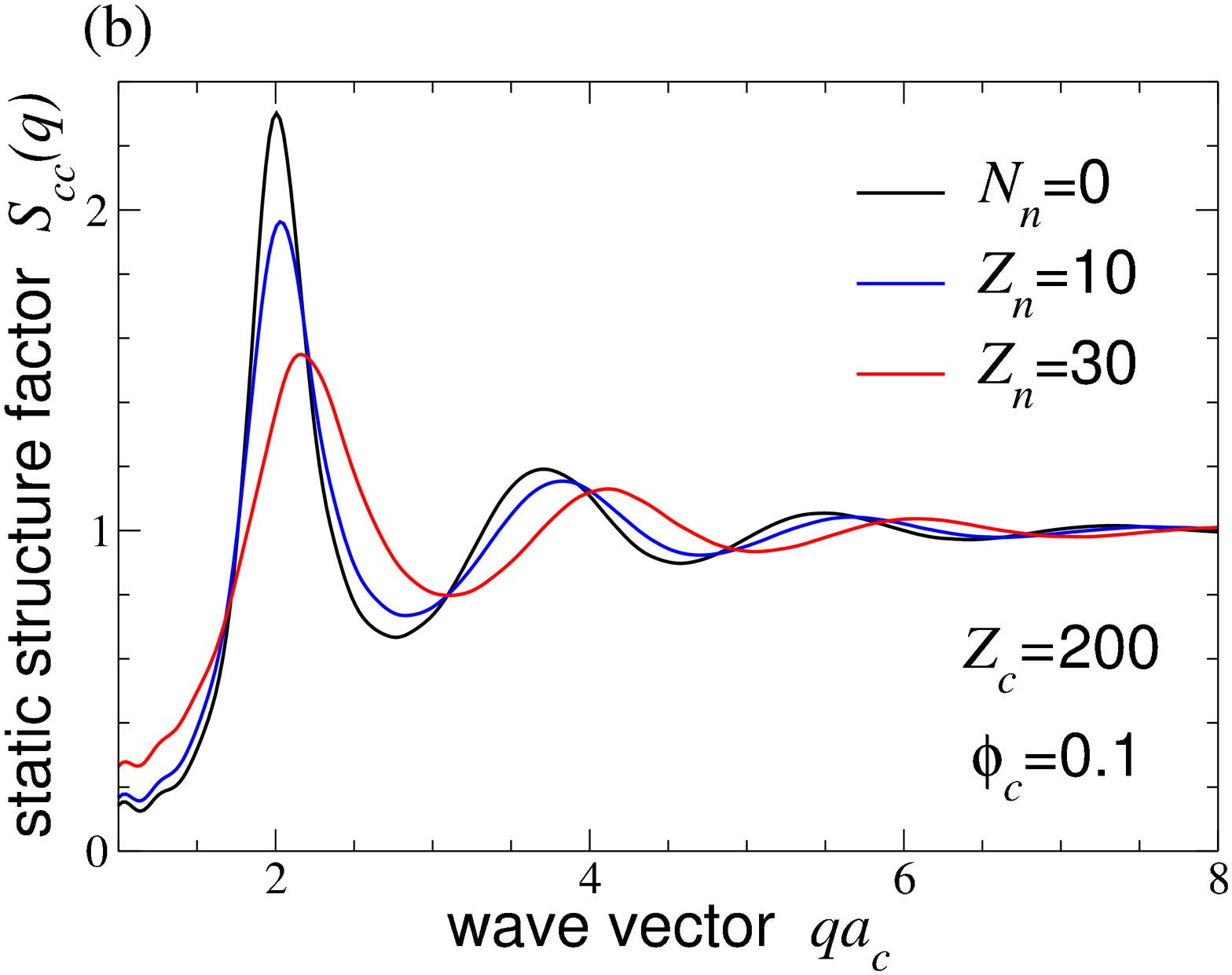}
\includegraphics[width=0.48\textwidth]{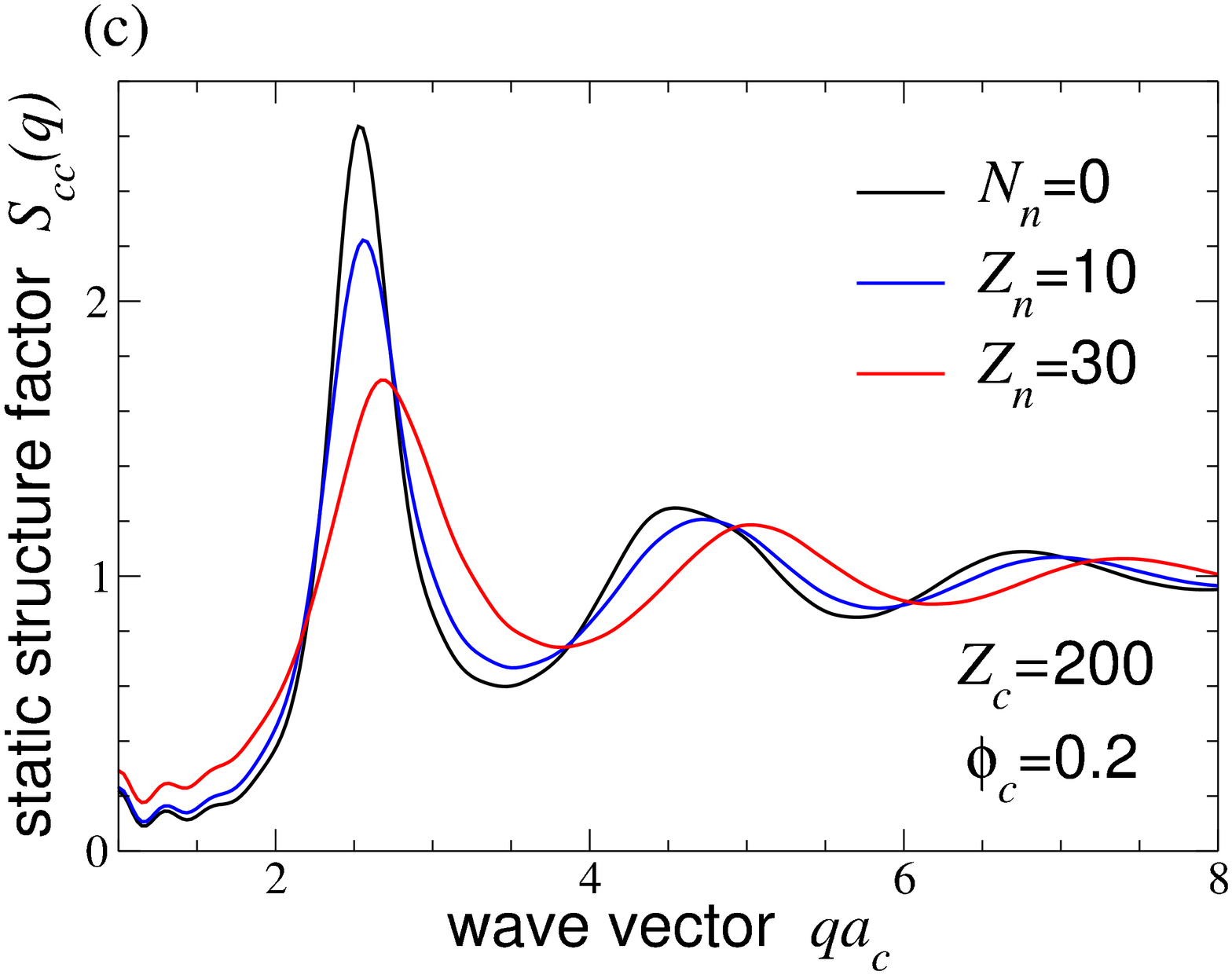}
\caption{Colloid-colloid static structure factors, scaled by a factor of $1/x_c$ 
[Eq.~(\ref{Sccq-scaled})], corresponding to radial distribution functions in Fig.~\ref{rdf1}
with colloid volume fractions $\phi_c=0.01$ (a), 0.1 (b), and 0.2 (c).
\label{ssf1}}
\end{figure}
 
\begin{figure}
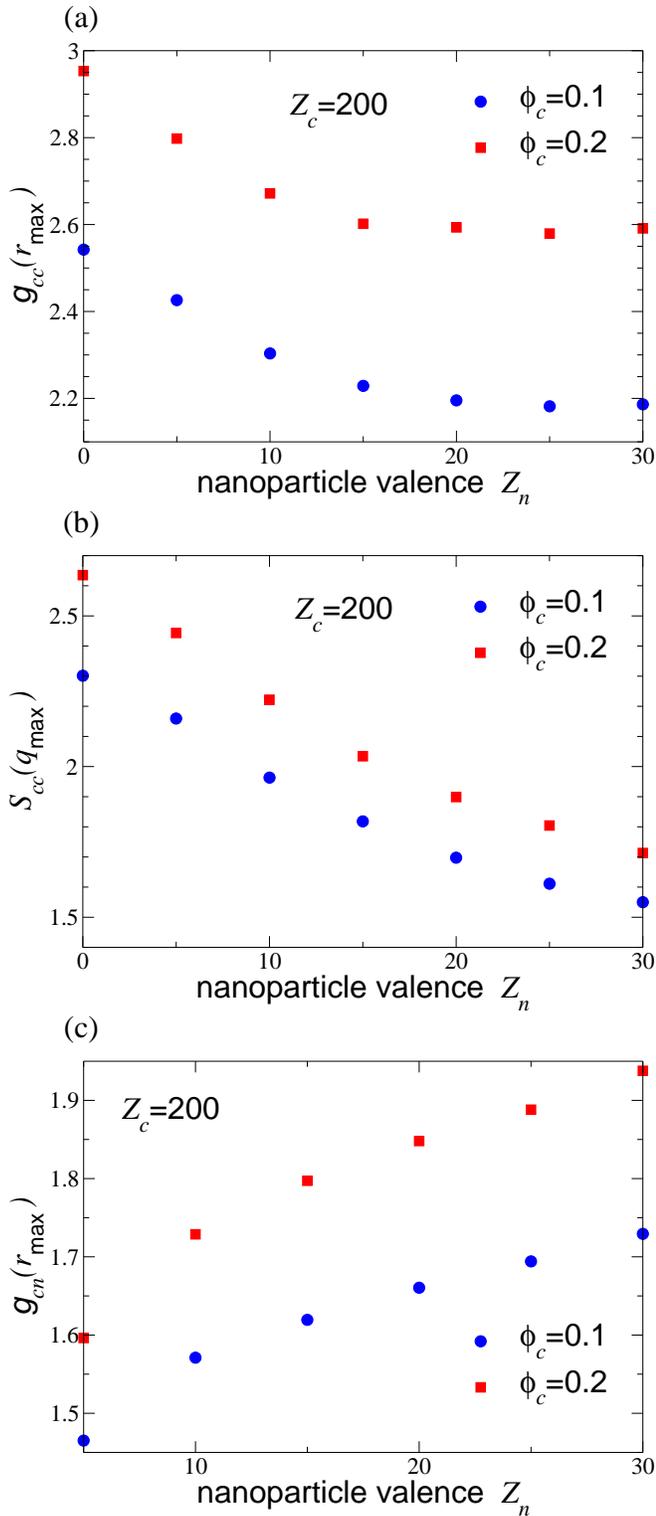

\includegraphics[width=0.48\textwidth]{grmax.nn1500.eps}
\includegraphics[width=0.48\textwidth]{sqmax.nn1500.eps}
\includegraphics[width=0.48\textwidth]{gcnrmax.nn1500.eps}
\caption{Height of main peak of (a) colloid-colloid radial distribution function (Fig.~\ref{rdf1}),
(b) colloid-colloid static structure factor (Fig.~\ref{ssf1}),
and (c) colloid-nanoparticle radial distribution function (Fig.~\ref{rdf1})
vs. nanoparticle valence $Z_n$ for
colloid number $N_c=500$, radius $a_c=50$ nm, and valence $Z_c=200$;
nanoparticle number $N_n=1500$ and radius $a_n=5$ nm;
and colloid volume fractions $\phi_c=0.1$ and 0.2.
\label{gr-sq-peak}}
\end{figure}

\begin{figure}
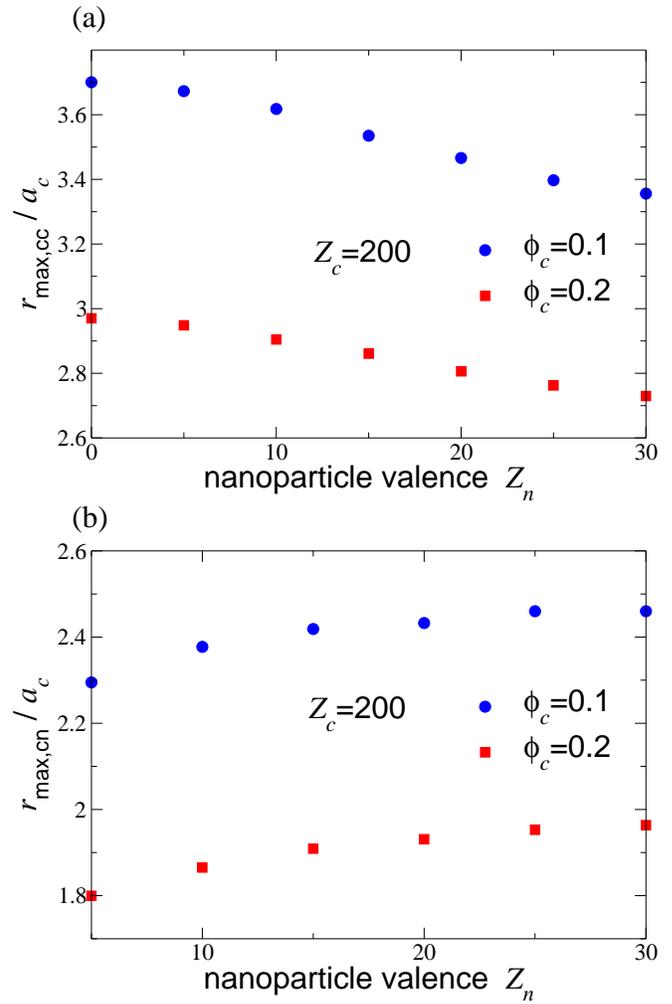

\includegraphics[width=0.48\textwidth]{rmaxgcc.nn1500.eps}
\includegraphics[width=0.48\textwidth]{rmaxgcn.nn1500.eps}
\caption{Position of main peak of (a) colloid-colloid and (b) colloid-nanoparticle 
radial distribution functions (Fig.~\ref{rdf1}) vs. nanoparticle valence $Z_n$ for
colloid number $N_c=500$, radius $a_c=50$ nm, and valence $Z_c=200$;
nanoparticle number $N_n=1500$ and radius $a_n=5$ nm;
and colloid volume fractions $\phi_c=0.1$ and 0.2.
\label{gcnr-peak-position}}
\end{figure}

\begin{figure}
\includegraphics[width=0.48\textwidth]{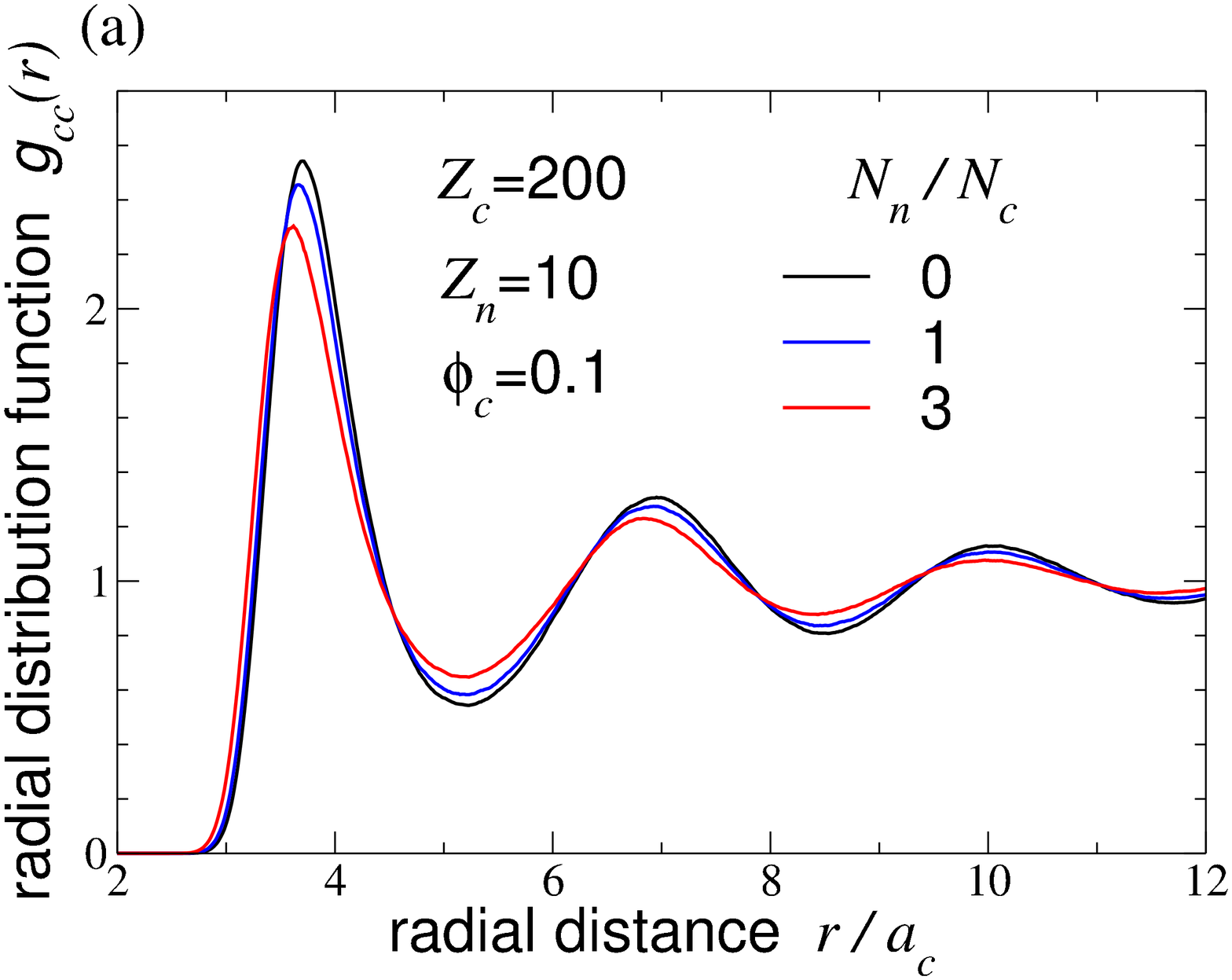}
\includegraphics[width=0.48\textwidth]{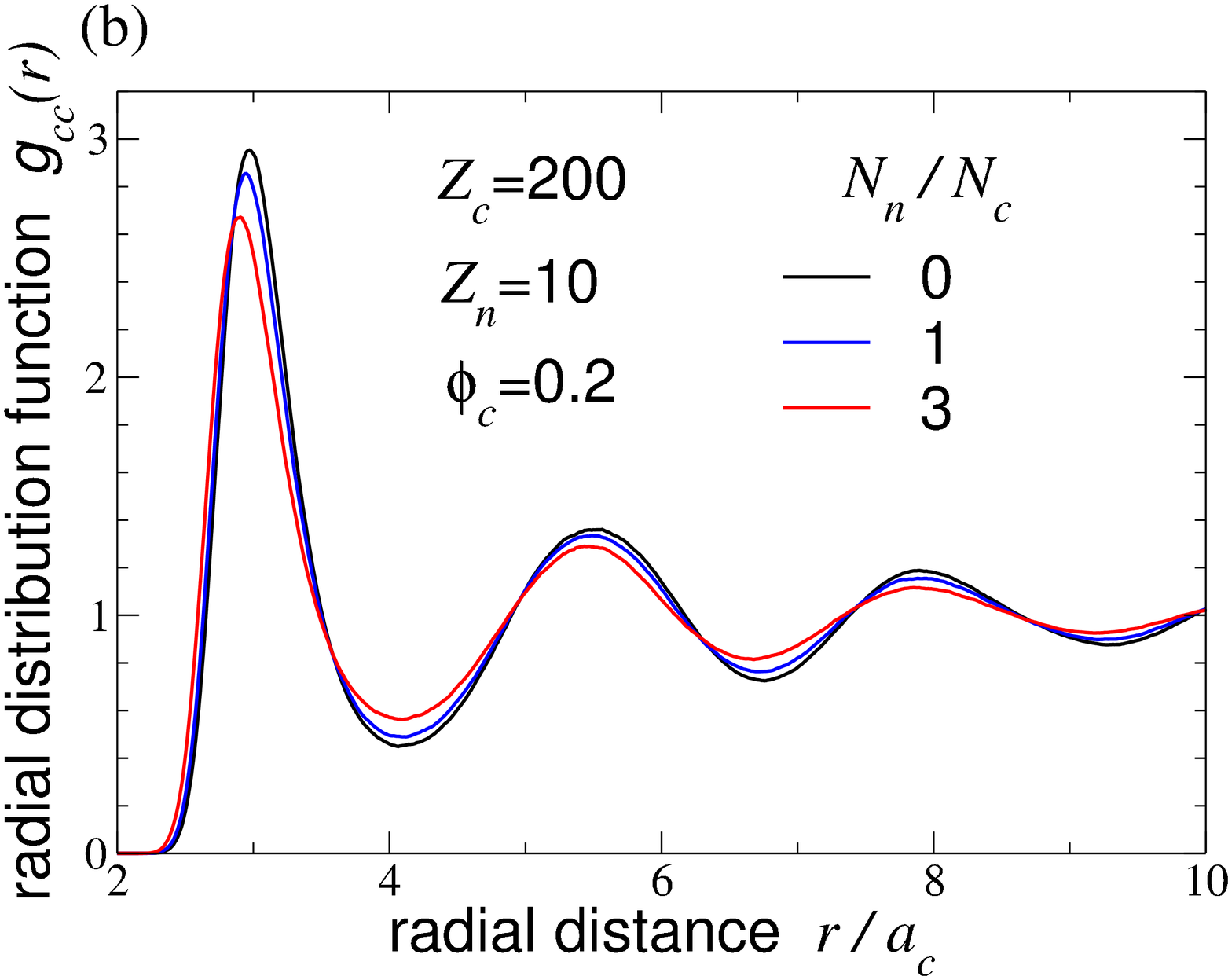}
\caption{Colloid-colloid radial distribution functions in salt-free mixtures of charged colloids 
and nanoparticles with colloid volume fractions $\phi_c=0.1$ (a) and 0.2 (b).
The colloid and nanoparticle valences are $Z_c=200$ and $Z_n=10$ and the number of 
nanoparticles per colloid, $N_n/N_c$, varies from 0 (black) to 1 (blue) to 3 (red).  
Other system parameters are the same as in Fig.~\ref{rdf1}.
With increasing nanoparticle concentration, colloid-colloid correlations become weaker,
as reflected by the decreasing height of the main peak of $g_{cc}(r)$.
\label{rdf2}}
\end{figure}

\begin{figure}
\includegraphics[width=0.48\textwidth]{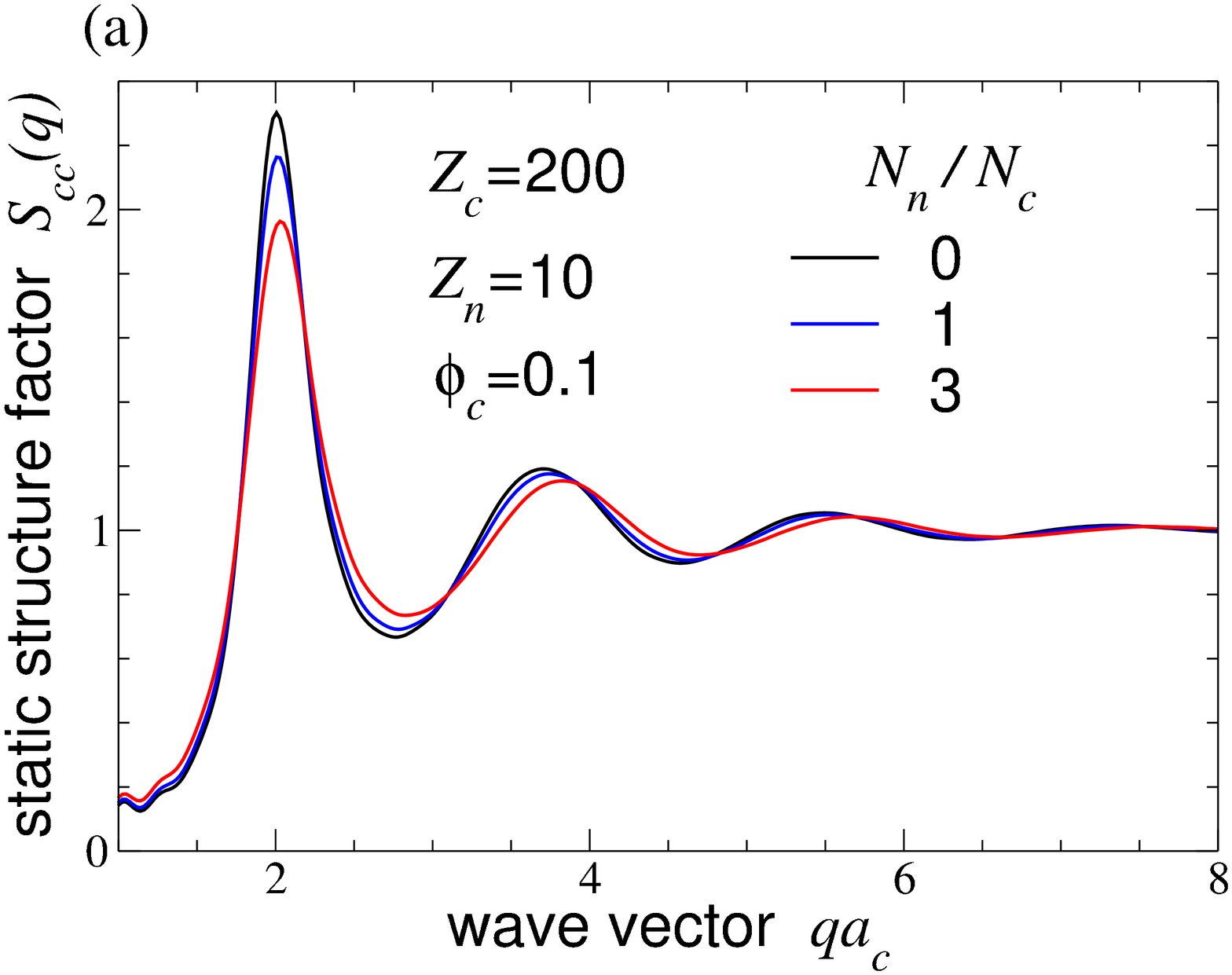}
\includegraphics[width=0.48\textwidth]{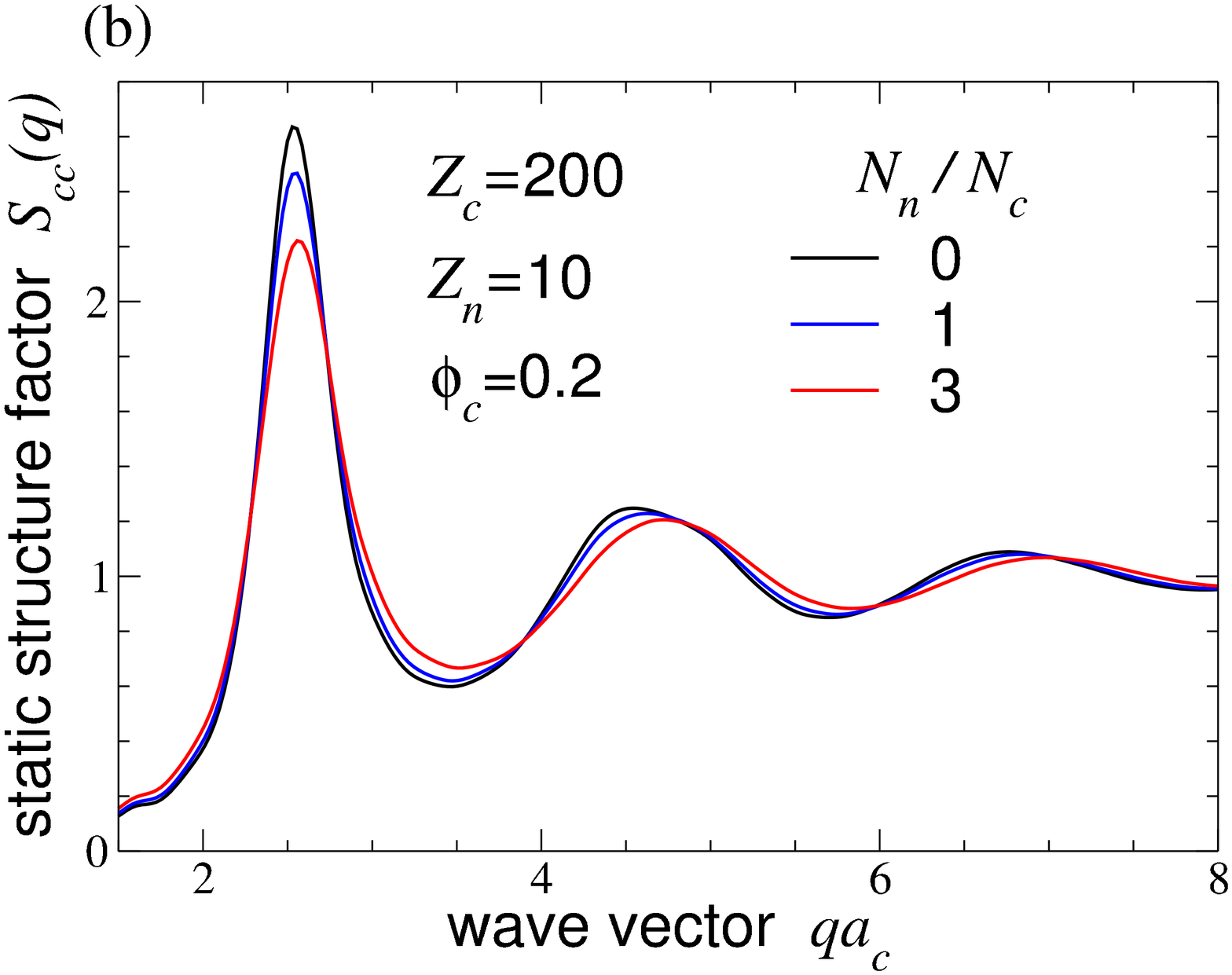}
\caption{Scaled colloid-colloid static structure factors corresponding to 
radial distribution functions in Fig.~\ref{rdf2} with colloid volume fractions 
$\phi_c=0.1$ (a) and 0.2 (b).
\label{ssf2}}
\end{figure}

\begin{figure}
\includegraphics[width=0.48\textwidth]{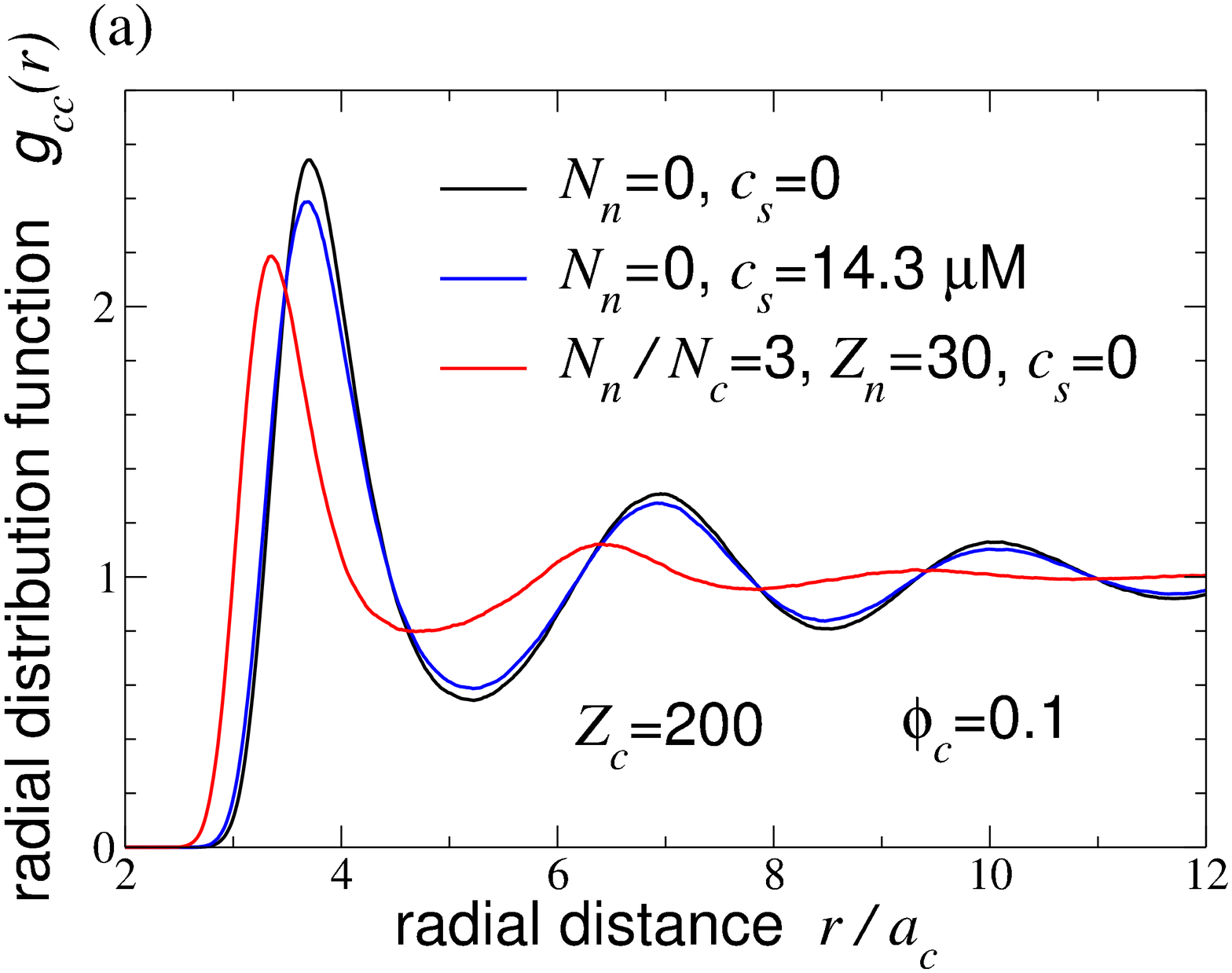}
\includegraphics[width=0.48\textwidth]{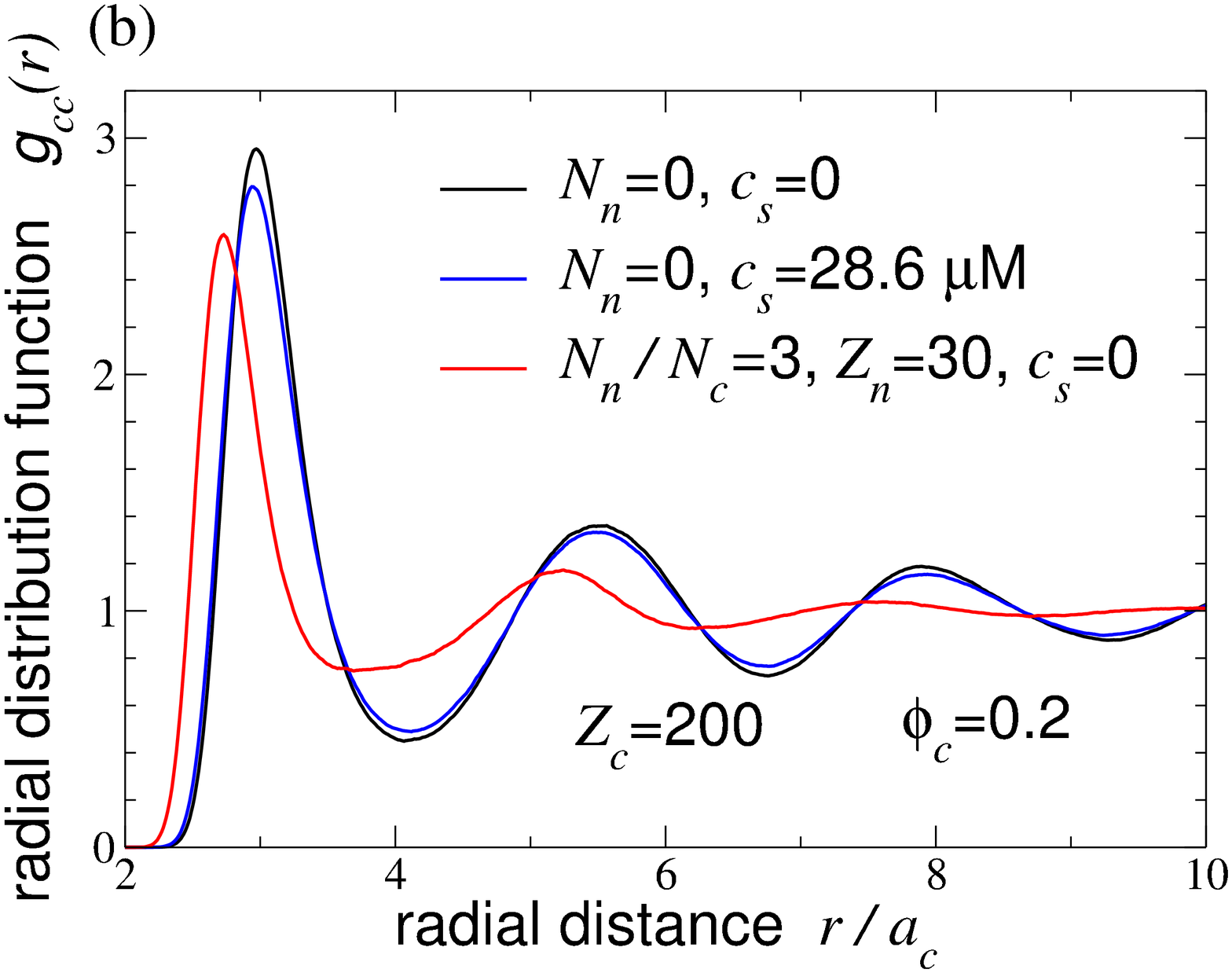}
\caption{Colloid-colloid radial distribution function in a colloid-nanoparticle mixture 
(red curves), with colloid and nanoparticle valences $Z_c=200$ and $Z_n=30$,
and in a nanoparticle-free suspension (blue curves) to which salt is added at 
concentration sufficient to yield the {\it same Debye screening constant}.
For reference, $g_{cc}(r)$ is shown also for the corresponding salt- and nanoparticle-free
suspensions (black curves).
For colloid volume fractions $\phi_c=$ 0.1 (a) and 0.2 (b), the required
salt concentrations are $c_s=14.3~\mu$M and $28.6~\mu$M.
Other system parameters are the same as in Fig.~\ref{rdf1}.
\label{rdf-salt}}
\end{figure}

\begin{figure}
\includegraphics[width=0.48\textwidth]{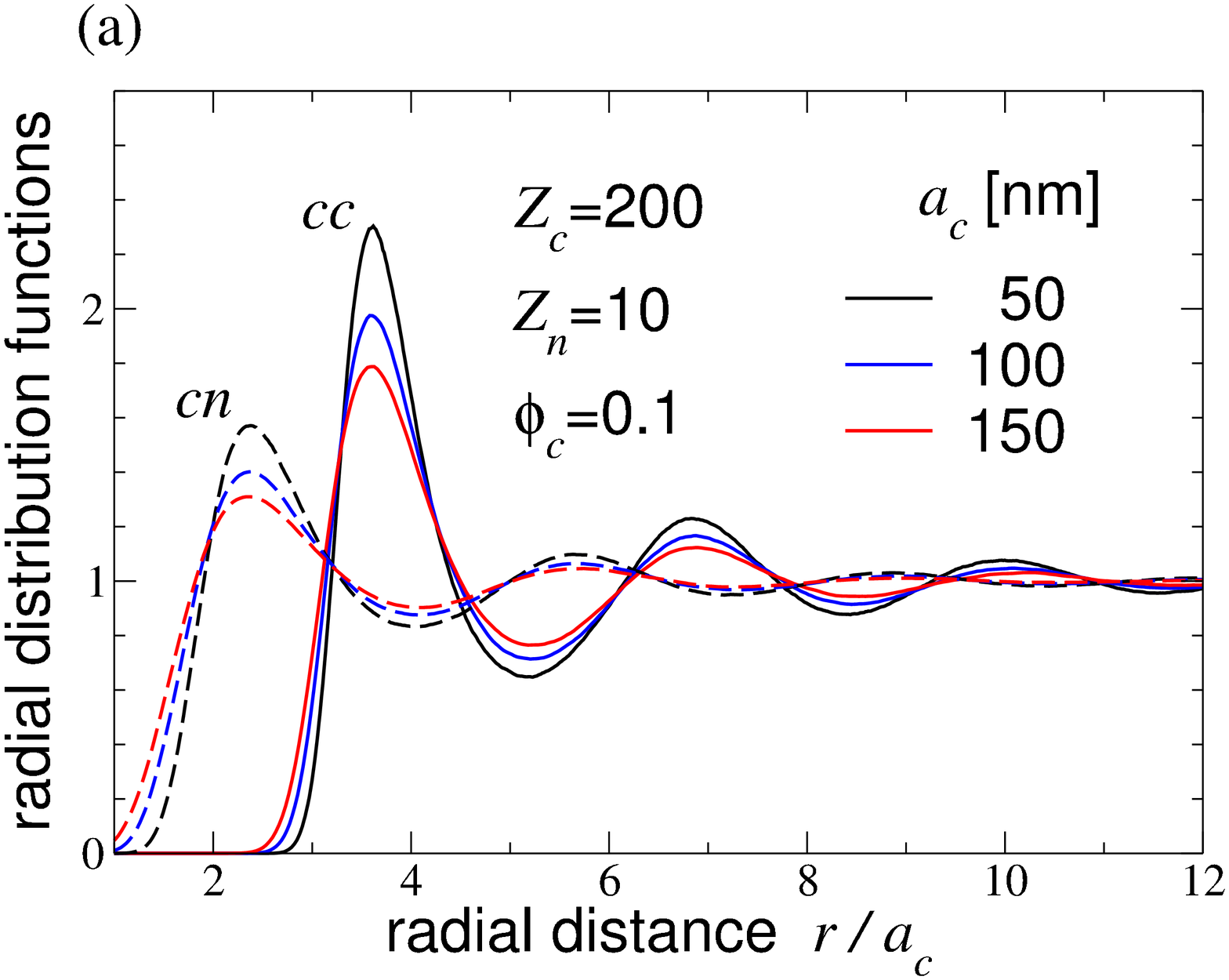}
\includegraphics[width=0.48\textwidth]{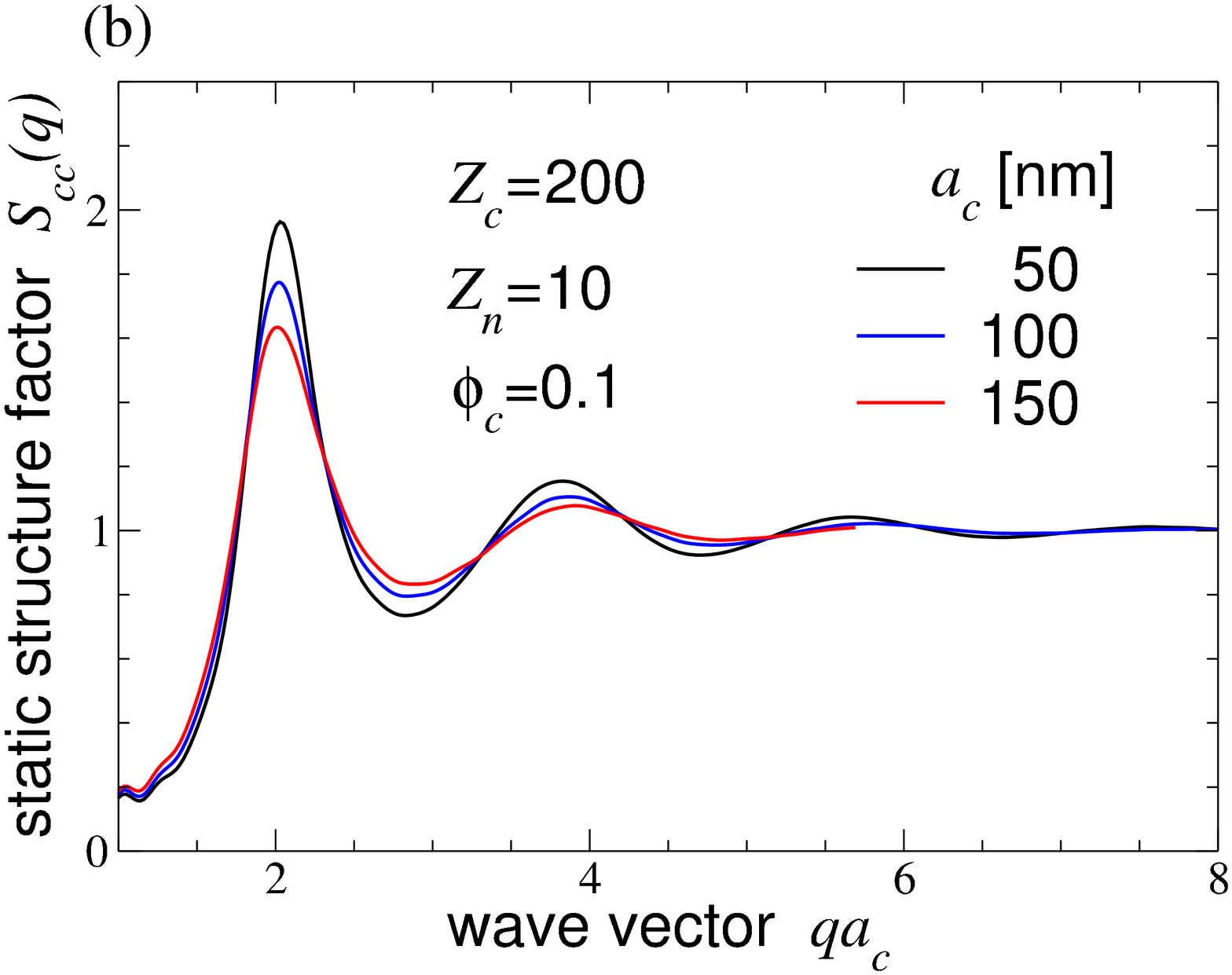}
\caption{Variation of colloid-colloid and colloid-nanoparticle radial 
distribution functions (a) and scaled colloid-colloid static structure factor (b) 
with colloid radius ($a_c=50, 100, 150$ nm) for fixed nanoparticle radius 
($a_n=5$ nm), valences ($Z_c=200$, $Z_n=10$), and colloid volume fraction ($\phi_c=$ 0.1).
In panel (a), solid and dashed curves, whose main peaks are labeled $cc$ and $cn$,
represent $g_{cc}(r)$ and $g_{cn}(r)$, respectively.
\label{rdf-sq-ac}}
\end{figure}

\begin{figure}
\includegraphics[width=0.48\textwidth]{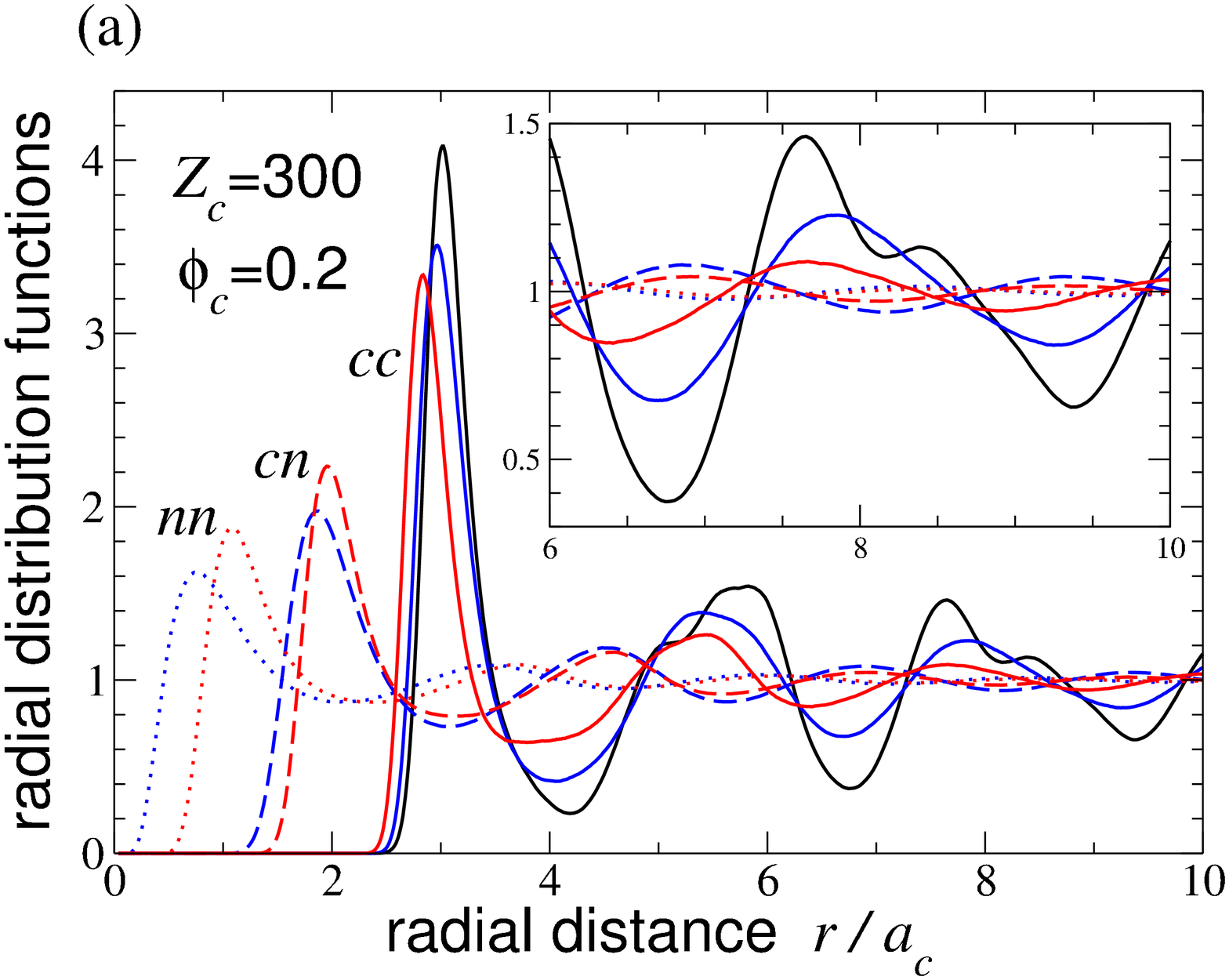}
\includegraphics[width=0.48\textwidth]{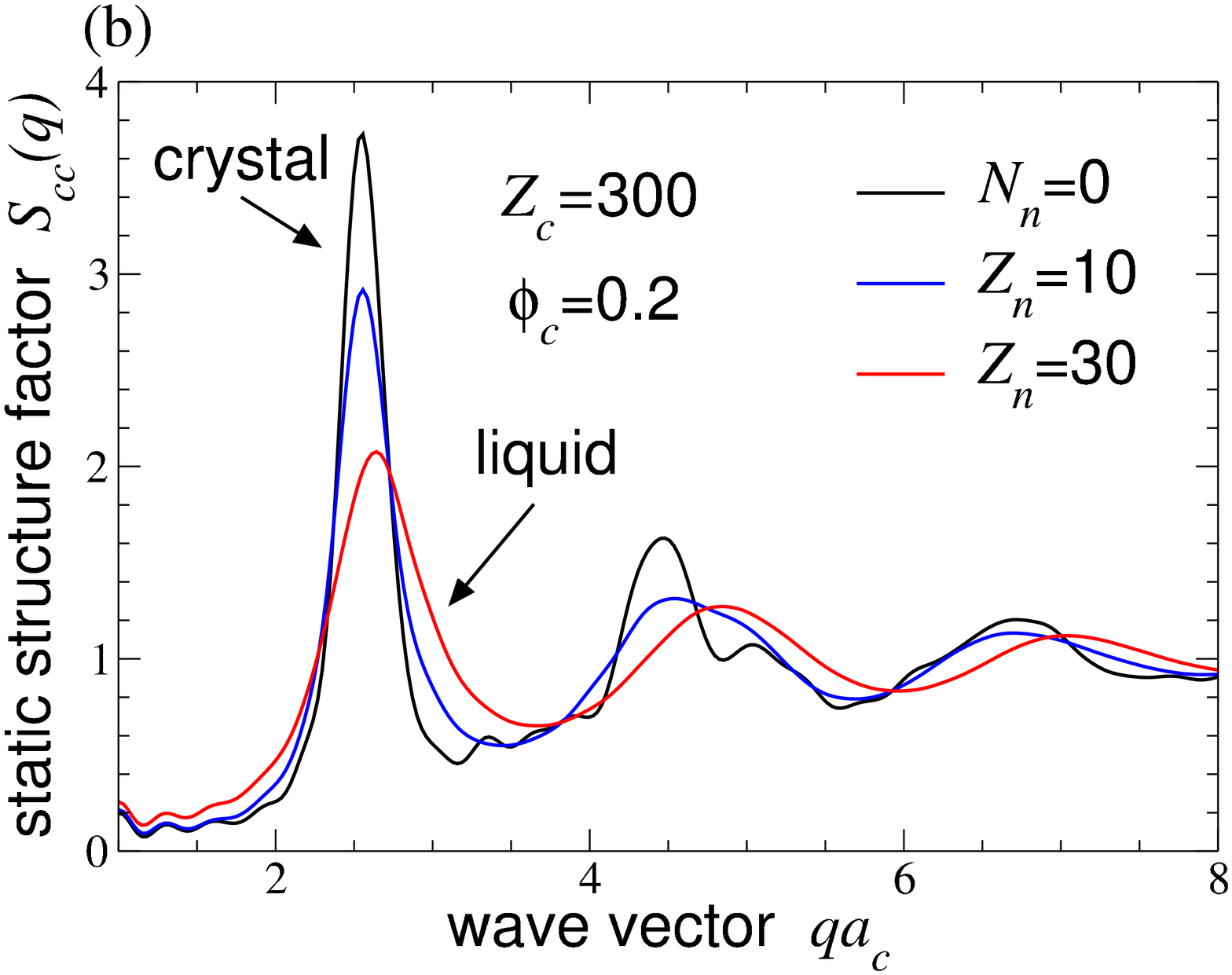}
\caption{(a) Radial distribution functions and (b) scaled colloid-colloid static 
structure factors for suspensions of $N_c=$ 500 colloids of valence $Z_c=300$ 
and volume fraction $\phi_c=0.2$, both without added nanoparticles (black) and 
with $N_n=1500$ nanoparticles of valence $Z_n=10$ (blue) or 30 (red).
In panel (a), solid, dashed, and dotted curves, whose main peaks are labelled 
$cc$, $cn$, and $nn$, represent $g_{cc}(r)$, $g_{cn}(r)$, and $g_{nn}(r)$, respectively. 
Inset magnifies long-ranged behavior.
Adding 3 nanoparticles per colloid of valence $Z_n=10$ weakens correlations (blue curves).
Increasing $Z_n$ to 30 further weakens correlations (red curves), inducing melting of 
the colloidal crystal, as reflected by the drop in height of the main peak of $S_{cc}(q)$.
\label{melting}}
\end{figure}

Figure~\ref{rdf1} presents our simulation data for radial distribution functions 
in mixtures with nanoparticle valences $Z_n=$ 10 and 30  and a concentration 
corresponding to 3 nanoparticles per colloid ($N_n=1500$).
Results are shown for colloid volume fractions $\phi_c=0.01$, 0.1, and 0.2. 
The Debye screening constants for these parameters range from $\kappa a_c=1$ to $\kappa a_c=2$.
For reference, results are also shown for $g_{cc}(r)$ from simulations of the 
one-component model of nanoparticle-free ($N_n=0$) colloidal suspensions.
Given the extreme size asymmetry ($a_n/a_c=0.1$), the total volume fractions 
of the nanoparticle-containing systems differ only slightly from $\phi_c$.
Upon addition of nanoparticles, the main peak of $g_{cc}(r)$ drops in height 
and shifts to shorter separation.  With increasing nanoparticle valence,
the peak drops further and shifts to yet shorter separations.

Complementary structural trends are observed in wave-vector space.  Figure~\ref{ssf1}
displays the corresponding colloid-colloid static structure factors, computed from 
particle configurations using Eq.~(\ref{Sccq}), and scaled by a factor of $1/x_c$ 
[Eq.~(\ref{Sccq-scaled})] for comparison with results for nanoparticle-free suspensions.
With increasing valence of added nanoparticles, the main peak of $S_{cc}(q)$ 
progressively drops in height and shifts to larger $q$.
These structural trends indicate that doping a charge-stabilized colloidal suspension 
with charged nanoparticles tends to weaken colloid-colloid correlations.

Also shown in Fig.~\ref{rdf1} are results for the colloid-nanoparticle and
nanoparticle-nanoparticle radial distribution functions, revealing the
nanoparticle correlations.  With increasing nanoparticle valence, the main peaks 
of both $g_{cn}(r)$ and $g_{nn}(r)$ rise in height and shift to longer separations.  
These variations are consistent with the stronger electrostatic repulsion between 
colloids and nanoparticles and between nanoparticle pairs with increasing $Z_n$.
Since $g_{cn}(r)$ is proportional to the probability of finding a nanoparticle 
a radial distance $r$ from the center a given colloid, a sharper main peak reveals 
a slight tendency for nanoparticles to accumulate near the colloids.  
Therefore, the colloids facilitate ordering of nanoparticles as well as
enhance correlations between nanoparticles.
Weakening of colloid-colloid correlations upon increasing nanoparticle charge
is thus accompanied by a modest strengthening of both colloid-nanoparticle and
nanoparticle-nanoparticle correlations.
It is important to emphasize that this gathering of nanoparticles around colloids,
which is driven by electrostatic repulsion, differs fundamentally from the 
nanoparticle adsorption and haloing observed in experiments~\cite{Lewis-2001-pnas,
Lewis-2001-langmuir,Lewis-2005-langmuir,chan-lewis2008-langmuir,Lewis-2008-langmuir}, 
which is associated with van der Waals attraction.

The structural trends exhibited in Figs.~\ref{rdf1} and \ref{ssf1} are 
systematically organized in Figs.~\ref{gr-sq-peak} and \ref{gcnr-peak-position}, 
which plot the main peak heights and positions of the radial distribution functions 
and static structure factors over a range of nanoparticle valence for two 
colloid volume fractions.  With increasing $Z_n$, the main peaks of $g_{cc}(r)$ 
and $S_{cc}(q)$ progressively drop in height, while the peak positions 
-- $r_{\rm max}$ for $g_{cc}(r)$ and $q_{\rm max}$ for $S_{cc}(q)$ -- shift to 
smaller $r$ and larger $q$, respectively.  At the same time, the main peaks of 
$g_{cn}(r)$ and $g_{nn}(r)$ monotonically increase in height and shift to larger $r$.
Secondary peaks show the same trends, consistent with weakening of colloid-colloid 
correlations and strengthening of colloid-nanoparticle correlations with increasing
nanoparticle valence.

Figures~\ref{rdf2} and \ref{ssf2} further illustrate the dependence of $g_{cc}(r)$ 
and $S_{cc}(q)$, respectively, on the concentration of nanoparticles, which varies 
from 0 to 3 nanoparticles per colloid.  The structural trends observed upon 
increasing $N_n$ at fixed $Z_n$ are qualitatively similar to those resulting from 
increasing $Z_n$ at fixed $N_n$.  Collectively, these results demonstrate that 
adding charged nanoparticles tends to weaken correlations between charged colloids.

The physical mechanism by which charged nanoparticles weaken correlations between 
colloids of like charge is subject to interpretation.  One possible perspective
is that a large charge asymmetry energetically favors configurations in which 
nanoparticles accumulate around the colloids, since the condition $|Z_n|\ll |Z_c|$ 
(or $|Z_c Z_n|\ll Z_c^2$) advantages colloid-nanoparticle interactions over 
colloid-colloid interactions.  The fact that the peak of $g_{cn}(r)$ lies closer 
to the colloid center for smaller $Z_n$ (greater charge asymmetry) and is higher 
and sharper for larger $Z_n$, when electrostatic coupling between colloids and
nanoparticles is stronger, may support this perspective.  However, the relative 
contributions of energy and entropy to the total free energy are difficult to assess 
and may depend on specific system parameters.  

Another conceivable interpretation is that charged nanoparticles may act to 
enhance the screening of electrostatic interactions between colloids.  To test 
this possibility, we performed additional simulations for colloidal suspensions
that are free of nanoparticles, but contain salt of sufficient concentration to
yield the same Debye screening constant as in a salt-free suspension with 
charged nanoparticles.  Figure~\ref{rdf-salt} compares the colloid-colloid 
radial distribution functions for these two systems.  Correlations between colloids
in the nanoparticle-containing, salt-free suspension are seen to be considerably 
weaker than in the salt-containing, nanoparticle-free suspension of the same
screening length.  The peaks of $g_{cc}(r)$ are lower and at shorter distances 
in the nanoparticle-containing suspension than in the salt-containing suspension.
This comparison demonstrates that ions confined to the surfaces of charged 
nanoparticles have a significantly stronger screening effect, and thus overall 
damping effect on correlations, than an equal number of free salt ions. 
Furthermore, this interpretation is consistent with conclusions drawn from an 
effective interaction theory, recently developed by one of us~\cite{Denton-pre2017}.
This theory, which is based on linear-response and random-phase approximations, 
maps a charged colloid-nanoparticle mixture onto a coarse-grained, one-component 
model of pseudo-colloids governed by an effective electrostatic pair potential 
\begin{equation}
v_{\rm eff}(r)\propto \frac{e^{-\kappa_{\rm eff}r}}{r},
\label{vcceff}
\end{equation}
involving a {\it nanoparticle-enhanced}, effective screening constant 
\begin{equation}
\kappa_{\rm eff}=\sqrt{\kappa^2+4\pi\lambda_B
\frac{\displaystyle Z_{n, \rm eff}^2 n_n}{\displaystyle 1-\phi'}}
\label{q}
\end{equation}
[compare with $\kappa$, defined in Eq.~(\ref{kappa})], where
\begin{equation}
Z_{n, \rm eff} = \frac{e^{\kappa a_n}}{1+\kappa a_n}Z_n
\label{Zneff}
\end{equation}
represents an effective valence of the nanoparticles, and $\phi'=(4\pi/3)n_c(a_c+a_n)^3$ is 
the volume fraction from which the nanoparticles are excluded by the colloids.
As confirmed by our simulations and the simulations of ref.~\cite{Denton-pre2017},
charged nanoparticles increase the effective screening constant much more efficiently 
than microions.  For example, a nanoparticle of valence $Z_n=10$ contributes more to
screening than 10 free microions.  We note that this interpretation differs somewhat 
from one recently suggested by Ojeda-Mendoza \etal~\cite{Moncho-Jorda-SM2018}, who 
attributed the influence of smaller particles in binary mixtures of charged colloids 
to an electrostatically-enhanced depletion attraction, albeit for size and charge 
asymmetries less extreme than considered here.

In addition to varying the charge asymmetry between colloids and nanoparticles,
we also performed additional simulations to explore the influence of size asymmetry
on structure.  Figure~\ref{rdf-sq-ac} shows radial distribution functions and 
static structure factors for three different colloid radii, $a_c=50$, 100, and 
150 nm, at fixed colloid volume fraction, $\phi_c=0.1$.  With increasing colloid size,
colloid-colloid and colloid-nanoparticle correlations become weaker, as signaled 
by the drop in height of the main peaks of $g_{cc}(r)$, $g_{cn}(r)$, and $S_{cc}(q)$.
This trend can be at least partially explained by the decrease in electrostatic 
coupling strength, $\lambda_B/a_c$, and the decrease in screening constant $\kappa$ 
[Eq.~(\ref{kappa})] that results from increasing $a_c$ at fixed $\phi_c$.
This behavior is opposite, however, what would be expected of uncharged colloids,
around which charged nanoparticles tend to accumulate to lower their potential energy,
the accumulation growing stronger the larger the volume displaced by the colloids.

We note in passing that, since the like-charged particles considered here interact via 
repulsive Yukawa forces that prevent their steeply repulsive cores from coming close to 
contact, conventional excluded-volume arguments regarding depletion of nanoparticles from 
the spaces between colloids do not cleanly apply here.  For this reason, the extent 
to which weakening of colloidal correlations can be directly associated with 
depletion-induced attraction is unclear and left open by our study.  
In the limit of uncharged particles, however, the system clearly reduces to 
an asymmetric mixture of hard spheres, for which density-functional theories of 
effective depletion-induced interactions have been developed~\cite{Dijkstra-pre1999,
Goetzelmann-epl1999,Roth-Evan-Dietrich-pre2000,Roth-jpcm2010}.  The crossover between 
a parameter regime dominated by electrostatic interactions and a regime dominated by 
excluded-volume interactions in charged hard-sphere mixtures is not well understood
and deserves future study.

A potential application of nanoparticle doping is illustrated in Fig.~\ref{melting}, 
which shows structural data for systems with a higher colloidal valence of $Z_c=$ 300.  
Suspensions of macroions of sufficient charge can self-assemble into ordered crystals.
According to the Hansen-Verlet freezing criterion~\cite{HansenVerlet1969}, 
the solid phase is stable when the height of the main peak of $S_{cc}(q)$ exceeds about 2.85.  
Correspondingly, $g_{cc}(r)$ exhibits pronounced peaks at lattice coordination distances.
When added to a colloidal crystal, nanoparticles may weaken 
colloidal interparticle correlations to the point of inducing melting of the crystal.
From inspection of Fig.~\ref{melting}, the nanoparticle-free suspension is 
clearly in a stable solid phase, with $S_{cc}(q_{\rm max})\simeq$ 3.8.  The positions
of the peaks of $g_{cc}(r)$ and $S_{cc}(q)$ are consistent with fcc crystal symmetry.
Upon adding nanoparticles in a 3:1 ratio ($N_n/N_c=3$), the colloidal correlations
are substantially weakened.  For nanoparticles of valence $Z_n=10$, the main peak height
of the structure factor drops to $S_{cc}(q_{\rm max})\simeq$ 2.9, placing the solid on the 
verge of melting.  The mixture of nondiffusing colloids and mobile nanoparticles is 
reminiscent of a superionic lattice (or sublattice melt) observed in atomic alloys.  
At higher nanoparticle valence ($Z_n=30$), colloidal correlations are furthered weakened 
and the structure factor clearly reveals that the crystal has melted.

\begin{figure}
\includegraphics[width=0.48\textwidth]{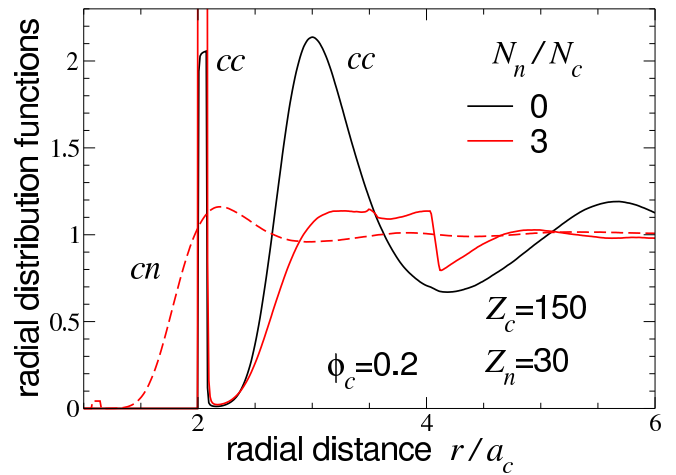}
\caption{Colloid-colloid ($cc$) and colloid-nanoparticle ($cn$) radial distribution functions 
(solid and dashed red curves, respectively) from our MD simulations of a salt-free mixture 
of like-charged colloids (valence $Z_c=150$, volume fraction $\phi_c=0.2$) and nanoparticles 
(valence $Z_n=30$), with three nanoparticles per colloid, interacting via effective pair potential 
of Fig.~\ref{vr2}.  
Also shown, for reference, is $g_{cc}(r)$ for the corresponding nanoparticle-free ($N_n=0$) 
suspension (solid black curve).  In the presence of van der Waals interactions, 
charged nanoparticles facilitate aggregation of weakly charged colloids,
as reflected in the sharp peak of $g_{cc}(r)$ near hard-core contact ($r=2a_c$)
and smaller secondary peaks.
\label{gr-vdw}}
\end{figure}
\begin{figure}[t]
(a)
\includegraphics[width=0.5\textwidth]{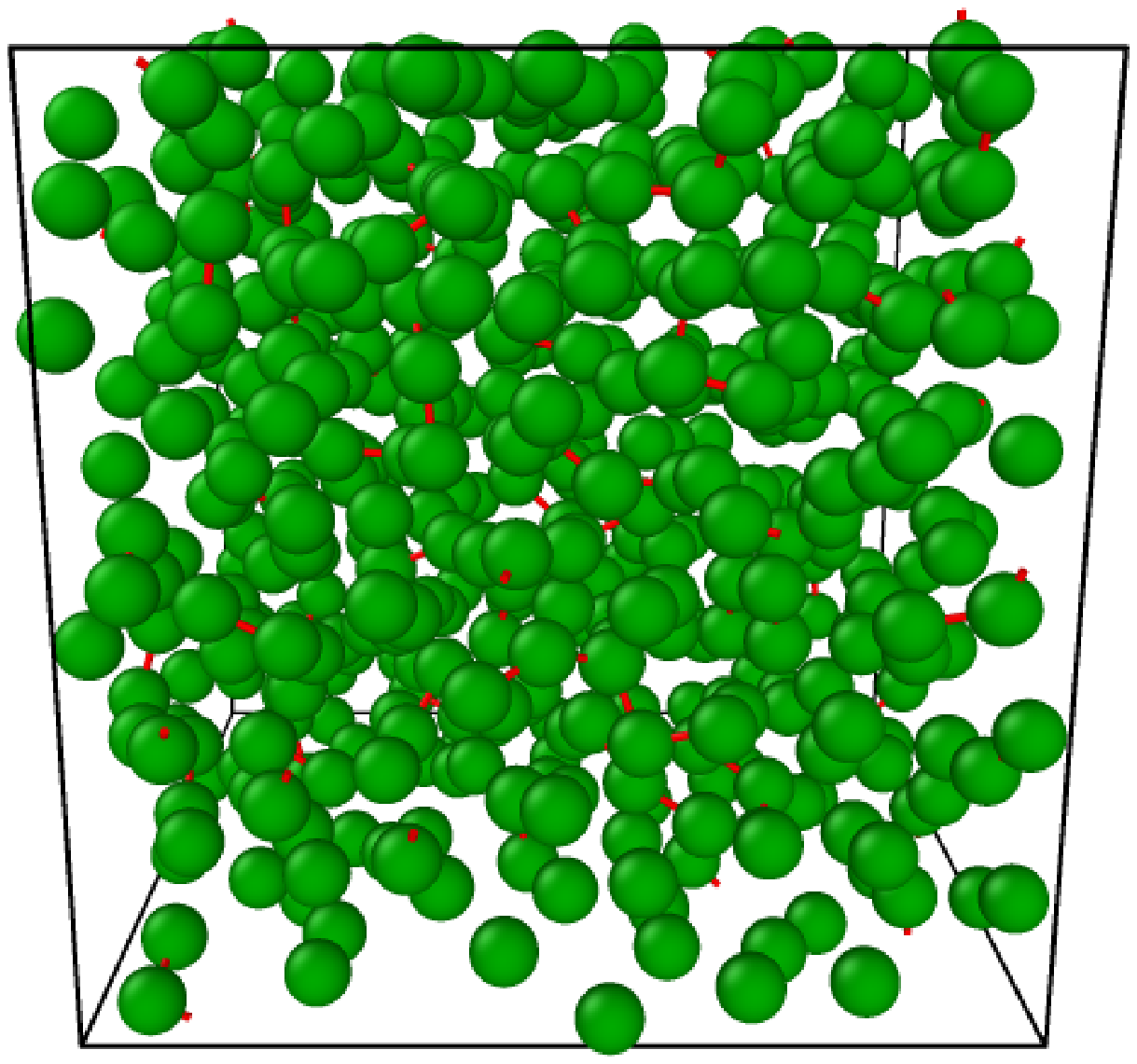}
\\
(b)
\includegraphics[width=0.5\textwidth]{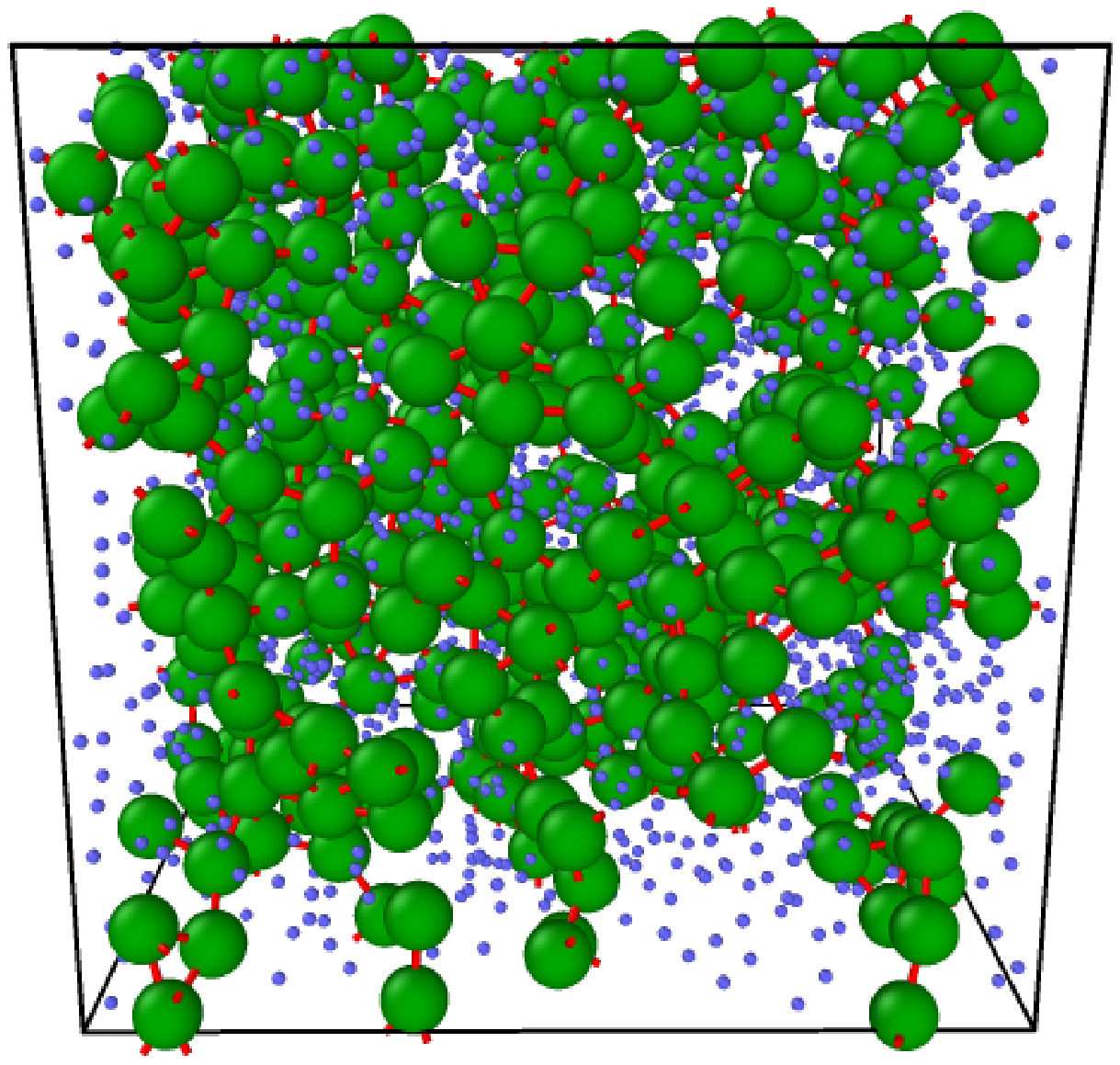}
\caption{Snapshots from simulations of (a) colloid-nanoparticle mixture and 
(b) pure (nanoparticle-free) colloidal suspension with particles of valences 
$Z_c=150$ and $Z_n=30$ interacting via electrostatic {\it and} van der Waals 
pair potentials.  Colloids (nanoparticles) are depicted as green (blue) spheres.  
Red bonds indicate pairs of aggregated particles.  To highlight bonds,
colloids are slightly reduced in size.
\label{snapshots-vdw}
}
\end{figure}

Finally, to elucidate the interplay between electrostatic and van der Waals interactions
in mixtures of like-charged colloids and nanoparticles, we performed additional simulations 
for systems interacting via the pair potentials shown in Fig.~\ref{vr2}.  Reducing the
colloid valence from $Z_c=200$ to 150 significantly lowers the barrier height of the total (DLVO) 
effective pair potential $v_{cc}(r)$, rendering bulk stability much more sensitive to the 
balance between competing repulsive electrostatic and attractive van der Waals interactions.
Figure~\ref{gr-vdw} shows representative results for radial distribution functions obtained
from runs of the same duration, but for $10^7$ steps, with the step now shortened to 1 fs,
as necessitated by the rapid spatial variations of the pair potentials.

In a pure, nanoparticle-free suspension, represented by the dashed curve in Fig.~\ref{gr-vdw},
the barrier of $v_{cc}(r)$ is sufficiently high ($\sim$20 $k_BT$) that the suspension is 
largely stable (or metastable) against van der Waals attraction-induced aggregation.  
Although a small fraction of colloid pairs have aggregated, as indicated by a narrow peak 
in $g_{cc}(r)$ near colloid-colloid contact ($r=2a_c$), most particles remain dispersed.
Adding nanoparticles significantly lowers the barrier height to $\sim$14 $k_BT$ and 
precipitates aggregation of colloids, as signaled by a sharp peak in $g_{cc}(r)$ at 
contact (solid $cc$ curve in Fig.~\ref{gr-vdw}).  Secondary peaks at larger separations 
reflect the onset of clustering.  Most prominent is the peak at $r=4a_c$, corresponding 
to three aggregated colloids in a linear configuration.
At the same time, electrostatic repulsion between colloids and nanoparticles gives a 
barrier height in $v_{cn}(r)$ of $\sim$16 $k_BT$ -- sufficiently strong to prevent adsorption 
of nanoparticles onto the colloid surfaces, as seen from the absence of a peak in $g_{cn}(r)$ 
at colloid-nanoparticle contact ($r=a_c+a_n$).  The nanoparticles, being free to screen 
electrostatic interactions, thus act to weaken colloid-colloid correlations and facilitate 
aggregation.  Figure \ref{snapshots-vdw} shows snapshots from these simulations in which
an aggregated pair of colloids is indicated by drawing a bond joining the two particles.
The figure illustrates the higher concentration of aggregated pairs in the mixture, 
as well as the dispersed distribution of nanoparticles.
It should be noted that, since colloidal aggregation is essentially irreversible,
the systems modeled in this case are not necessarily in thermodynamic equilibrium.
Nevertheless, our results clearly show the destabilizing influence of charged nanoparticles.

\section{Conclusions}\label{Conclusions}
In summary, by performing a computer simulation study of a coarse-grained model
of charged colloid-nanoparticle mixtures, in which particles interact via 
physically consistent effective pair potentials, we have explored the influence 
of nanoparticles on the structure and stability of colloidal suspensions.
In contrast with most previous experimental and modeling work, our study focuses 
on mixtures of like-charged (mutually repulsive) particles, in which the
colloids have sufficiently high valences that repulsive electrostatic interactions 
usually dominate over attractive van der Waals interactions.
While our results for like-charged colloid-nanoparticle mixtures are not
directly comparable with experimental observations reported for unlike-charged 
mixtures~\cite{Lewis-2001-pnas,Lewis-2001-langmuir,Lewis-2005-langmuir,
chan-lewis2008-langmuir,Lewis-2008-langmuir}, they qualitatively corroborate
previous conclusions that charged nanoparticles can significantly affect the 
stability of colloidal suspensions.

By computing radial distribution functions and static structure factors,
we have shown that addition of charged nanoparticles to a charge-stabilized
colloidal suspension can significantly weaken correlations between colloids.
Furthermore, with increasing charge of nanoparticles, while colloid-colloid 
correlations grow weaker, colloid-nanoparticle correlations grow stronger.
The latter trend is consistent with amplification of the electrostatic coupling 
with increasing particle charge.  The weakening of colloid-colloid correlations 
can be interpreted as an enhancement of electrostatic screening 
by charged nanoparticles, as predicted by a theory of effective electrostatic 
interactions in charged colloid-nanoparticle mixtures~\cite{Denton-pre2017}.  
Our results demonstrate that nanoparticle-enhanced screening tends to promote 
irreversible aggregation of weakly charged colloidal particles in a fluid suspension 
and can induce melting of colloidal crystals.

The accuracy of the coarse-grained model studied here, and the validity of our physical 
interpretations of the influence of charged nanoparticles on the structure of charge-stabilized 
colloidal suspensions, may be checked by more extensive simulations of the primitive model,
which includes microions explicitly~\cite{louis-allahyarov-loewen-roth2002,Luijten2004,
Linse2005,Dijkstra2007,Dijkstra2010}.
Future studies may also extend our parameter survey by varying particle sizes, charges, 
and concentrations and exploring oppositely-charged and multicomponent mixtures.  
The interplay between electrostatic and van der Waals interactions deserves
further exploration, especially across the transition from systems dominated by 
electrostatic interactions to those dominated by excluded-volume interactions.
Also important would be analyzing the influence of nanoparticles on thermodynamic 
properties (e.g., osmotic pressure) and phase transitions, such as freezing/melting.
Such an analysis would require consistently incorporating the volume energy and 
accounting for the density dependence of the effective pair potentials~\cite{Denton2010}. 
In the end, our study may shed light on the possibilities of exploiting nanoparticles
to tune effective interactions and guide self-assembly of colloidal materials.

\vspace*{0.2cm}
\begin{acknowledgments}
This work was supported by the National Science Foundation under Grant No.~DMR-1106331.
We thank Dr.~Jun Kyung Chung and Dr.~Hartmut L\"owen for helpful discussions 
and acknowledge the Center for Computationally Assisted Science and Technology 
at North Dakota State University for computing resources.
\end{acknowledgments}


%

\end{document}